\documentclass[11pt]{article}
\usepackage[utf8]{inputenc}
\usepackage{graphicx,pstricks,pst-node,psfrag,amsthm,amssymb,amsmath,dsfont,bbm}
\usepackage[authoryear]{natbib}
\usepackage[a4paper]{geometry}
\usepackage{setspace}
\usepackage{float} %for begin{figure}[H]
\usepackage{subcaption} %subfigure
\usepackage[labelfont=bf]{caption}
\usepackage{array}
\usepackage{booktabs}
\usepackage{authblk}
\usepackage{tikz}
\usetikzlibrary{decorations.pathreplacing,calc}
\usetikzlibrary{positioning}
\usepackage{multirow}
\AtBeginEnvironment{tabular}{\sffamily}
\AtBeginEnvironment{figure}{\sffamily}
\AtBeginEnvironment{tikzpicture}{\sffamily}

\newcommand{\Var}{\operatorname{Var}}
\newcommand{\Cor}{\operatorname{Cor}}
\allowdisplaybreaks[1]

\begin{document}

\title{Excessive data censoring in fMRI undermines individual precision and weakens brain-behavior associations}
\author{Amanda Mejia, Joanne Hwang, Damon Pham, Stephanie Noble, Theodore D. Satterthwaite, Thomas E. Nichols, B.T. Thomas Yeo} 

\date{}

\maketitle

\doublespacing

\begin{abstract}
Censoring high-motion volumes in functional MRI is a common practice to reduce the effects of head motion in functional connectivity (FC) analyses. Although aggressive censoring may remove more noise, it causes extensive data loss, creating a tradeoff that may ultimately improve or degrade FC accuracy. Here, we evaluate how motion censoring affects FC estimation and downstream brain-wide association studies (BWAS). Using extensively sampled participants from the Human Connectome Project (HCP) Retest dataset, we establish individual ``ground truth'' FC and assessed the accuracy of FC estimated from 5 to 30 minute scans. We find that censoring consistently degrades FC accuracy, with more aggressive censoring being more detrimental, particularly among participants exhibiting above-average motion levels. In these participants, aggressive censoring reduces FC accuracy by 30\% for 30-minute scans denoised with ICA-FIX, an advanced noise removal method, and by 3\% for scans denoised with conventional confound regression. These effects reflect substantial data loss (34\% reduction in effective scan duration) that outweighs comparatively modest noise reductions: 7\% with ICA-FIX and 18\% with confound regression. Compensating for scrubbing-induced FC accuracy loss would require substantially longer scans (62\% with confound regression; 76\% with ICA-FIX), inflating data collection budgets. Introducing a repeated measures framework to separate motion trait from artifact, we find that standard quality control metrics are dominated by motion trait and overstate motion bias, which is effectively mitigated with less aggressive censoring. Finally, using data from nearly 1,000 participants in the main HCP, we demonstrate that unreliable FC substantially attenuates BWAS correlations: by $\sim$30\% under optimal conditions (longer ICA-FIX scans with no censoring) but exceeding 75\% in short, aggressively censored scans. Our findings support the use of ICA-FIX or other advanced denoising methods over confound regression, avoiding or limiting censoring, and collecting at least 15 minutes of data to maximize fidelity of individual-level FC and BWAS.
\end{abstract}

\section{Introduction}

Participant head motion during functional magnetic resonance imaging (fMRI) is a major source of noise and artifact, which can systematically bias functional connectivity (FC) \citep{satterthwaiteImpactInScannerHead2012, vandijkInfluenceHeadMotion2012, powerSpuriousSystematicCorrelations2012}. A common mitigation strategy is censoring volumes concurrent with or proximal to motion \citep{powerSpuriousSystematicCorrelations2012, power2015recent}. Censoring has been shown to reduce motion artifacts \citep{ciricBenchmarkingParticipantlevelConfound2017, parkesEvaluationEfficacyReliability2018, satterthwaiteMotionArtifactStudies2019} and is common practice in fMRI research \citep{Marek2022, lynch2024frontostriatal, ooi2025longer}.

However, whether censoring is necessary on top of modern fMRI denoising techniques is a matter of debate \citep{pruim2015evaluation, parkesEvaluationEfficacyReliability2018, reddy2024denoising}. Answering this question is urgent, because censoring has drawbacks that may undermine large, costly fMRI data collection efforts and hinder neuroscience research and clinical care. Censoring can substantially shorten scan duration \citep{pruim2015evaluation, pham2023less}, but longer scans are important for reliable FC \citep{laumann2015functional} and are cost-effective for brain-wide association studies (BWAS) \citep{ooi2025longer}. Exclusion of participants due to insufficient data after censoring creates a less representative sample and induce selection bias, since motion is associated with socioeconomic status, ethnicity, cognitive ability, and disease severity \citep{nebel2022accounting}. High exclusion rates also make the huge samples needed for reliable BWAS \citep{Marek2022} more elusive. In the ABCD study analyzed by \cite{Marek2022}, more than 60\% of over 10,000 participants were excluded due to censoring. Censoring also reduces degrees of freedom, lowering FC reliability \citep{parkesEvaluationEfficacyReliability2018}. This attenuates BWAS correlations and inflates their variance, leading to systematic underestimation and poor reliability of already weak effects.

The impacts of censoring on FC and BWAS depend on the choice of censoring approach, which varies widely. The most common form of censoring is based on participant head motion and signal intensity change, using motion thresholds from lenient to stringent. Some pipelines additionally remove neighboring volumes without evident abnormalities. A common ``expanded'' strategy excludes one preceding and two subsequent volumes around high-motion frames \citep{powerSpuriousSystematicCorrelations2012} and discards segments with fewer than five contiguous volumes \citep{powerMethodsDetectCharacterize2014}. This expansion can remove many unaffected volumes, leading to substantial data loss. These different approaches to censoring may have drastically different effects on FC and BWAS. Thus, the question is not only whether to censor, but how aggressively, and according to which metric(s).

In this study, we comprehensively evaluate the effects of motion censoring on two primary and complementary use cases of fMRI. First, we test its impact on individual-level FC estimation, which is critical for biomarkers and clinical translation. Second, we assess downstream implications for BWAS, a central tool in population neuroscience.

For individual-level analyses, we study $38$ highly sampled participants from the Human Connectome Project (HCP) Retest dataset. For each, we define ``ground truth'' FC using 1.5 hours of stringently censored fMRI data and compare it to FC estimated from data censored to varying degrees and of varying duration (5 to 30 minutes). We find that censoring consistently worsens FC estimates relative to ground truth, particularly with expanded censoring in higher-motion individuals. We show that censoring reduces underlying noise, but this is outweighed by drastic data loss, ultimately degrading FC accuracy. To achieve comparable accuracy, expanded censoring requires approximately 50\% longer scans than lenient censoring.

These findings motivate us to re-examine how motion bias in FC is traditionally assessed. Much of the rationale for aggressive censoring derives from a quality control measure, QC-FC, which quantifies motion–FC associations \citep{power2015recent, ciricBenchmarkingParticipantlevelConfound2017, parkesEvaluationEfficacyReliability2018}. However, the conventional implementation of QC-FC using cross-sectional data conflates transient motion artifact with stable individual differences associated with motion tendency \citep{zeng2014neurobiological}. To address this limitation, we extend the QC-FC framework to repeated measures data using the main HCP sample ($N=1087$), allowing us to distinguish motion artifact from genuine trait-like variation. We find that standard QC-FC is dominated by motion trait and overstates the presence of motion bias, while repeated-measures QC-FC shows that less aggressive censoring strategies effectively mitigate artifact.

Finally, we examine how censoring and scan duration influence the strength of BWAS effect sizes. It is recognized that large samples are required to estimate stable brain–behavior correlations \citep{Marek2022}, but the reliability of brain and behavioral measures are equally critical \citep{noble2021guide, nikolaidis2022suboptimal, gell2023burden}. Imperfect measurement reliability attenuates correlations, so improving the fidelity of measures underlying BWAS is important to maximize effect sizes \citep{tiego2022putting}. Using $N = 990$ HCP participants with sufficient data to define ground-truth FC, we show that expanded censoring and shorter scans markedly weaken BWAS correlations by reducing FC reliability. Across realistic scenarios, attenuation ranges from $\sim$30\% under favorable conditions (longer ICA-FIX-processed scans with no censoring, highly reliable behavioral measure) to over 75\% with short scans and expanded censoring. Thus, preprocessing and acquisition choices can impose substantial, often unrecognized limits on the strength of brain–behavior associations that can be recovered from fMRI data.

\section{Results}

We examine the effects of censoring in fMRI on the validity of downstream participant-level analysis, motion-related noise, data collection requirements, and BWAS. We consider a wide range of censoring approaches: no censoring \citep[e.g.,][]{zalesky2014time, zhao2022common, segal2023regional},  
censoring using lenient and stringent motion thresholds \citep[e.g., ][]{Marek2022, lynch2024frontostriatal}, and expanded censoring \citep[e.g.,][]{powerMethodsDetectCharacterize2014, ooi2025longer}. Expanded censoring is defined as exclusion of one preceding and two subsequent volumes for each stringently censored volume, as well as removal of remaining segments containing fewer than five contiguous volumes (Fig. \ref{fig:FD_Comparison}). Volumes are censored if they exhibit high motion based on either framewise displacement (FD) or DVARS (Derivative of VARiance over Spatial locations), a data-driven measure of head motion.  We use a filtered and lagged version of FD appropriate for the multiband fMRI data in the Human Connectome Project \citep{power2019distinctions} as implemented in \cite{pham2023less}. We adopt FD thresholds of 0.2mm (stringent) to 0.5mm (lenient). For DVARS we adopt a statistically principled adaptive threshold \citep{afyouni2018insight}. 

Because the potential benefit of censoring depends on other data denoising techniques applied to the data, we consider two different baseline denoising methods representing opposite ends of the spectrum: regression of 36 motion and global signal parameters (36P) \citep{fristonMovementrelatedEffectsFMRI1996, satterthwaiteImprovedFrameworkConfound2013} and ICA-FIX, a data decomposition approach \citep{griffanti2014ica}. Previous research has found nuisance regression to insufficiently remove motion artifacts \citep{kim2025evaluating}, while ICA-based denoising has been shown more effective \citep{pruim2015evaluation}, especially when combined with global signal regression \citep{ciricBenchmarkingParticipantlevelConfound2017, parkesEvaluationEfficacyReliability2018}. Since ICA-FIX as implemented in the HCP does not include global signal regression (GSR), here we consider a version of ICA-FIX with GSR to facilitate comparison with 36P denoising. See Methods Section \ref{methods:scrub_and_denoise} for details of data processing and censoring.

\subsection{More aggressive censoring degrades FC accuracy}
\label{sec:FC_validity}

% FIGURE 1 (SECTION 2.1): Worse MAE with stringent censoring

\begin{figure}
    \centering
    \includegraphics[width=0.9\linewidth]{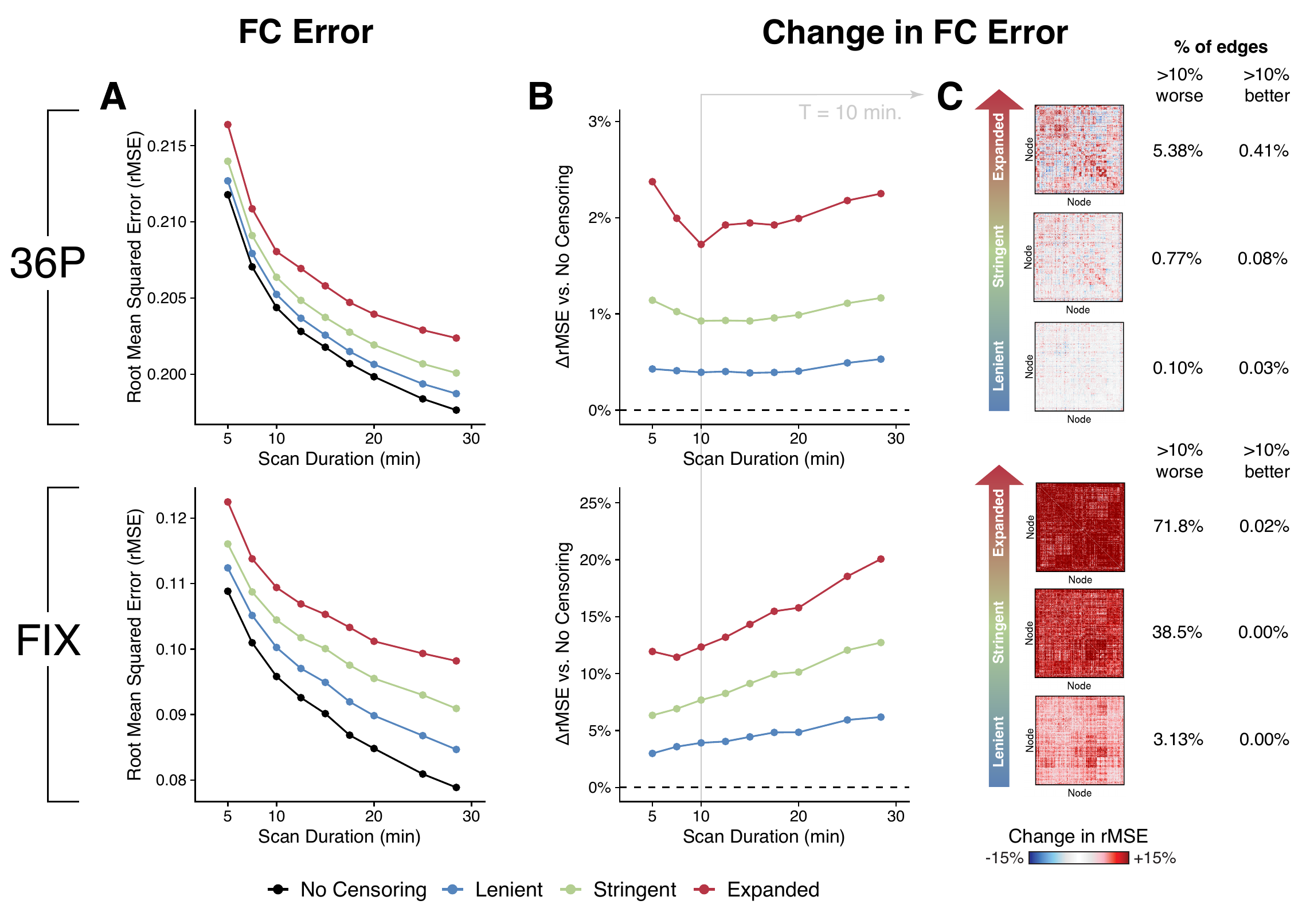}
    \caption{\small \textbf{More aggressive censoring worsens accuracy of individual FC estimates.} FC error is relative to individuals' ``ground truth'' based on 90 minutes of stringently censored data per participant.  Root mean squared error (rMSE) is over participants and edges in panels A and B and over participants for each edge in panel C. Panels B and C show change in rMSE compared to no censoring. \textbf{(A)} FC error decreases with longer acquired scan duration, but more censoring leads to progressively higher error across all durations in both 36P-processed and FIX-processed data. FC error is much lower overall in FIX-processed data than in 36P-processed data.
    \textbf{(B)} Longer scans do not mitigate the detrimental effects of censoring on FC accuracy; rather, the gap between censoring and no censoring widens with increasing scan duration. This reflects an increase in the censoring rate as the scan progresses, due to higher participant motion levels later in the scan. Censoring is particularly harmful in FIX-processed data, where expanded censoring results in over 20\% higher FC error for 30-minute scans.
    \textbf{(C)} Across edges, the effects of censoring are mixed for 36P-processed data, with most edges worsening but some improving ($T=10$ minutes). While the effect of expanded censoring in 36P data is mild on average across all edges, many edges show much stronger detrimental effects (indicated in dark red), particularly within-network edges and subcortical-cortical connections. Expanded censoring worsens FC error by at least $10\%$ in over $5\%$ of edges, while it improves FC error by the same amount in only $0.4\%$ of edges.  For FIX-processed data, by contrast, censoring universally worsens FC accuracy across the connectome: over 71\% of edges worsen by at least 10\%, while effectively zero edges improve by the same amount.
    }

    \label{fig:FC_accuracy}
\end{figure}

To assess how censoring affects the accuracy of individual-level functional connectivity (FC) estimates, we first established a “ground truth” FC matrix for each of 38 participants in the HCP retest dataset \citep{van2013wu} with sufficient data (Methods Section \ref{methods:FC}). Ground truth FC was computed from approximately 90 minutes of stringently censored data per participant. FC is defined as the pairwise Pearson correlation among 419 cortical surface and subcortical regions \citep{schaeferLocalGlobalParcellationHuman2018, fischl2012freesurfer} (see Figure \ref{fig:region_labels}). We then estimate FC from the remaining 30 minutes of data while varying both scan duration (5–30 minutes) and censoring strategy (none, lenient, stringent, expanded). FC error is quantified as the root mean squared error (rMSE) between estimated and ground truth FC, averaged over participants and/or edges. 

Across both preprocessing pipelines (36P and ICA-FIX), FC error decreased monotonically with longer scan duration (Figure \ref{fig:FC_accuracy}A), consistent with improved reliability from additional data. However, censoring consistently worsened FC accuracy at every duration, and more aggressive censoring was more detrimental. The FC accuracy achieved with expanded censoring of 30-minute scans was similar to the FC accuracy of $\sim$10-15 minute scans without censoring. Lenient censoring produced relatively modest increases in rMSE ($\approx 0.5\%$ with 36P or $\approx 5\%$ with FIX), while stringent and especially expanded censoring led to progressively larger increases in error (Figure \ref{fig:FC_accuracy}B). Notably, longer scans did not mitigate these detrimental effects. Instead, the gap between no censoring and expanded censoring widened with increasing duration. This surprising result reflects an increase in the censoring rate as the scan progresses, due to participants moving more later in the scan. 

The detrimental impact of censoring was substantially larger in FIX-processed data. Even lenient censoring increased FC error by several percent, and expanded censoring produced marked degradation across all durations (Fig. \ref{fig:FC_accuracy}B). For 30-minute scans, expanded censoring resulted in over 20\% higher error relative to no censoring. Note that these results reflect averages over all edges, and some edges may experience much worse degredation, as we now examine.

Edge-wise analyses further revealed that censoring effects were heterogeneous in 36P-processed data but nearly universally detrimental in FIX-processed data (Figure \ref{fig:FC_accuracy}C). For 36P, most edges showed worsening with expanded censoring at 10 minutes, though a small minority improved. Expanded censoring worsened FC error by $>10\%$ in $5.4\%$ of edges, whereas only $0.4\%$ of edges improved by the same magnitude. Stronger detrimental effects were observed particularly for within-network connections and subcortical–cortical edges. By contrast, FIX-processed data exhibited consistent worsening across the connectome: 71.8\% of edges showed $>10\%$ increases in error, and virtually none showed comparable improvement.

Across all scan durations, higher FC error with more aggressive censoring was found to be statistically significant in both 36P- and FIX-processed data. We used paired Wilcoxon rank sum tests to compare the participant-level rMSE over edges at different levels of censoring. Lenient censoring was significantly worse than no censoring; stringent censoring was significantly worse than lenient censoring; and expanded censoring was significantly worse than stringent censoring, with $p < 0.001$ after Bonferroni correction (Figure \ref{fig:tests_alldurations}). 

\subsection{Excessive data loss outweighs noise reduction, especially for high movers}
\label{sec:worse_for_high_movers}

% FIGURE 2: FC error stratified by motion level + effective scan duration & baseline noise

\begin{figure}
\centering
\includegraphics[width=0.9\linewidth]{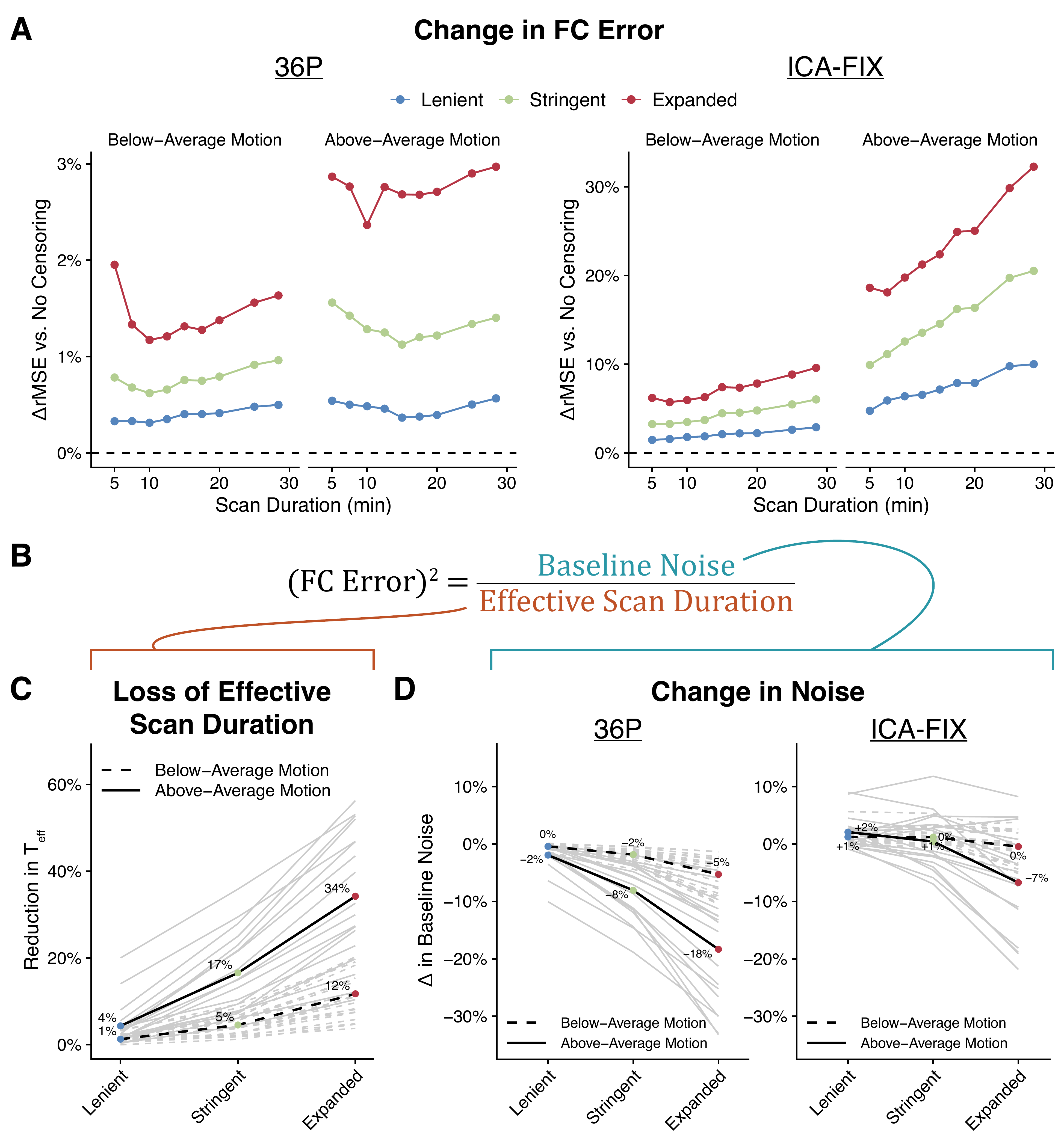}
\caption{\small \textbf{Censoring is more detrimental for participants who move more, due to excessive data loss}. \textbf{(A)} For participants with above-average motion levels (see Figure ~\ref{fig:motion_split}), censoring increases FC error more than in low-motion participants. For FIX-processed data, the effects of expanded censoring are dramatic in these participants, worsening FC error by over 30\% for 30-minute scans. \textbf{(B)} FC error is determined by the ratio of two competing factors: baseline noise and scan duration, both of which are typically decreased by censoring. \textbf{(C)} There is a dramatic loss in effective scan duration due to expanded censoring, particularly in high-motion participants, with data loss of 34\% on average. \textbf{(D)} More aggressive censoring reduces baseline noise, but not as much as it reduces effective scan duration. For high-motion participants, expanded censoring reduces baseline noise by 14\% on average in 36P-processed data and by 5\% in FIX-processed data. One outlying participant with approximately 40\% higher baseline noise in FIX-processed data is excluded for visualization purposes and from the average to avoid its undue influence. In low-motion participants, the reduction in baseline noise is mild in 36P-processed data and is non-existent in FIX-processed data. Across motion levels, any reductions in baseline noise is outweighed by more dramatic data loss, ultimately leading to higher FC error.}
\label{fig:FC_accuracy_bymotion}
\end{figure}

We next stratified participants into below-average and above-average movers (Figure \ref{fig:motion_split}) to examine whether the effects of censoring on FC accuracy differed as a function of motion level. Several high-motion participants in the HCP Retest dataset were already excluded due to insufficient data for ground truth FC estimation (Methods Section \ref{methods:ground_truth}); thus, the “above-average” group represents the upper half of motion within a fairly low-motion sample.

In both motion groups, censoring increased FC error relative to no censoring, but the magnitude of this effect was larger in above-average movers (Figure \ref{fig:FC_accuracy_bymotion}A). In 36P-processed data, expanded censoring increased FC error by approximately 1.5\% in below-average movers and by approximately 3\% in above-average movers, with some variation depending on scan duration. In FIX–processed data, effects were substantially larger. In below-average movers, expanded censoring increased FC error by roughly 6–10\%,  depending on duration, while in above-average movers it increased FC error by approximately 20–30\%, reaching over 30\% for 30-minute scans. Lenient and stringent censoring produced smaller increases in error, but these effects were also consistently larger in above-average movers and in FIX–processed data.

To examine the basis of these differences, we decomposed FC error into two components: baseline noise and effective scan duration (Methods Section \ref{methods:baseline_var}). Because squared error is equal to baseline noise divided by effective scan duration (Figure \ref{fig:FC_accuracy_bymotion}B), censoring can influence FC accuracy by reducing noise (leading to lower error), reducing effective duration (leading to higher error), or both. Thus, the overall impact of censoring on FC accuracy will depend on whether the degree of data loss is outweighed by noise reduction, or vice-versa. 

Expanded censoring led to substantial reductions in effective scan duration, especially in above-average movers. Expanded censoring reduced effective duration by approximately 34\% in above-average movers on average and by 12\% in below-average movers in both 36P-processed data (Figure \ref{fig:FC_accuracy_bymotion}C) and FIX-processed data (results not shown). Stringent censoring reduced effective duration by approximately 17\% and 5\% in these groups, respectively, while lenient censoring resulted in only minimal reductions.  

Across motion groups and preprocessing pipelines, these reductions in effective scan duration were consistently greater than the corresponding reductions in baseline noise (Figure \ref{fig:FC_accuracy_bymotion}D). On average, expanded censoring reduced baseline noise in 36P-processed data by 18\% in above-average movers and by 5\% in below-average movers. In FIX–processed data, reductions were much smaller: expanded censoring reduced baseline noise by 7\% in above-average movers and showed no reduction in below-average movers. These reductions in baseline noise were outweighed by the corresponding reductions in effective scan duration, ultimately leading to higher FC error.  These results also elucidate that the much greater detrimental effect of censoring on FC accuracy in FIX-processed data is due to negligible reductions in baseline noise, accompanied by substantial data loss. 

\subsection{More aggressive censoring necessitates longer acquisitions, leading to larger budgets and/or fewer participants}
\label{sec:budget_inflation}

%Figure 3: Required Scan Duration

\setlength{\tabcolsep}{1pt}
\begin{figure}
    \centering
    \includegraphics[width=\textwidth]{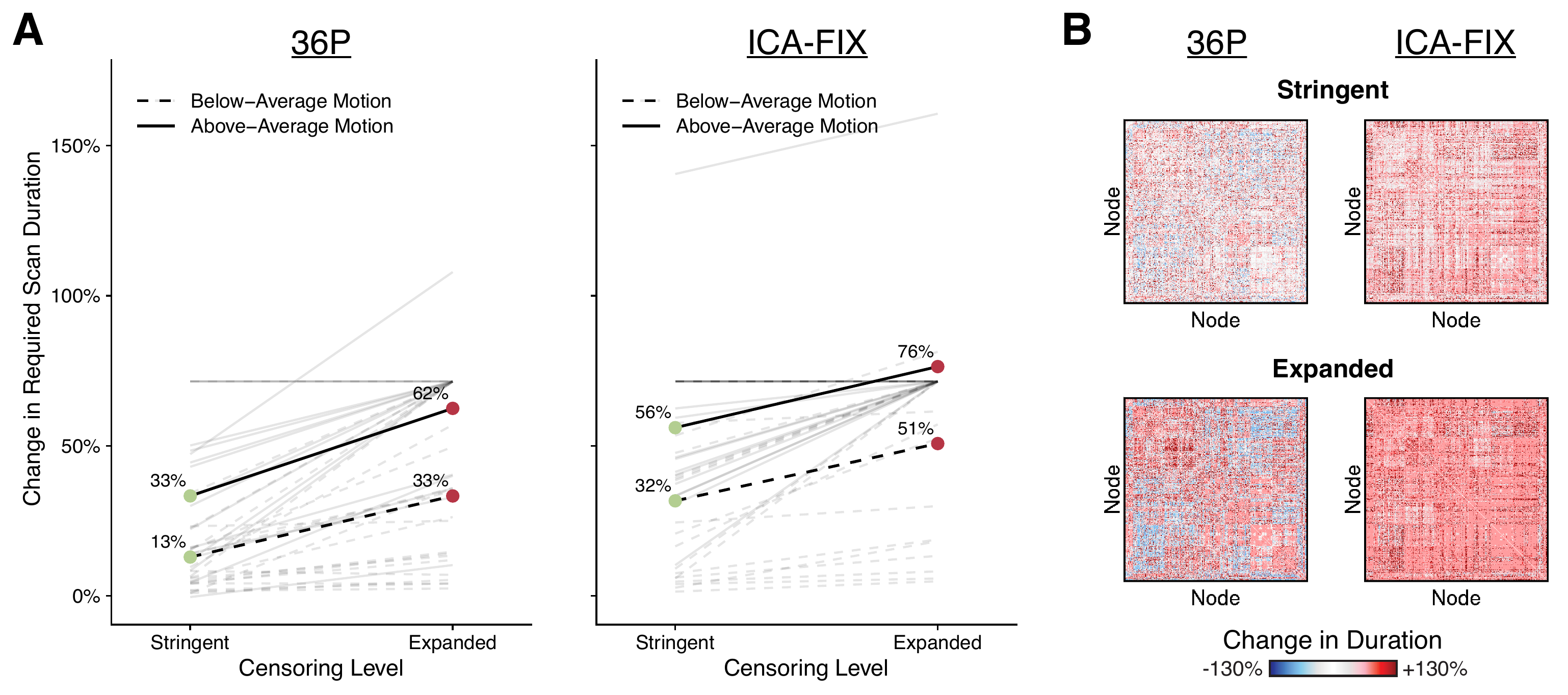}
    \caption{\small \textbf{Stringent and expanded censoring necessitate substantially longer scans to maintain FC accuracy.} Values represent the percentage change in scan duration required to maintain edge-wise FC accuracy when more stringent motion censoring is used, compared with lenient censoring (see Figure \ref{fig:durChange_illustration} for illustration). \textbf{(A)} Percent change in required scan duration to maintain FC accuracy compared with lenient censoring. Mean over edges for each participant is shown, along with the mean over participants within each motion group. For high-motion participants, expanded censoring requires over 60\% longer scan durations in 36P-processed data and over 75\% longer scans in FIX-processed data on average. \textbf{(B)} Edge-wise percent change in required scan duration, averaged over participants. Color scale is truncated at $\pm 130\%$, but values range from -77\% to +500\%. Scan duration would need to at least double to maintain FC accuracy with expanded censoring for 13.5\% of edges in 36P-processed data and 15.1\% in FIX-processed data.}
    \label{fig:durationChange}
\end{figure}

Because more aggressive censoring increases FC error through disproportionate loss of effective scan duration compared to gains in noise reduction, we next quantified how much additional data acquisition would be required to maintain FC accuracy relative to lenient censoring. For each participant and for each edge, we determined the percent increase in scan duration required to match the FC accuracy achieved with lenient censoring of a 17.5-minute scan (Methods Section \ref{methods:required_scan_duration}; Figure \ref{fig:durChange_illustration}). The need for longer scans would result in larger data collection budgets and/or reduced numbers of participants.

More stringent censoring required longer acquisitions in both preprocessing pipelines and motion groups to maintain FC accuracy on average over all edges (Figure \ref{fig:durationChange}A). In 36P-processed data, stringent censoring required, on average, 33\% longer scans in above-average movers and 13\% longer scans in below-average movers. Expanded censoring required approximately 62\% longer scans in below-average movers and 33\% longer scans in above-average movers. In FIX–processed data, the required increases were substantially larger, given the more drastic detrimental effects of censoring on FC accuracy after denoising with ICA-FIX. Stringent censoring required 56\% longer scans in above-average movers and 32\% longer scans in below-average movers. Expanded censoring required approximately 76\% longer scans in above-average movers and 51\% longer scans in below-average movers.  

Edge-wise results (Figure \ref{fig:durationChange}B) showed that most edges required longer acquisitions under more aggressive censoring in 36P-processed data, while in FIX-processed data nearly all edges required longer acquisitions. For both processing pipelines, there was substantial variability over edges, with some edges requiring much greater increases in scan duration compared with the average.  In particular, expanded censoring would necessitate at least doubling scan duration to maintain FC accuracy for 13.5\% of edges in 36P-processed data and for 15.1\% of edges in ICA-FIX–processed data. 

Note that these results correspond to the average over participants. Since scan duration requirements generally increase more for above-average movers, for some participants the percent of edges that require at least doubling scan duration will be much higher. Note also that, since scan duration cannot in practice be increased longer for some edges than others, the observed heterogeneity over edges has an important implication:  if the entire connectome, or certain disproportionately affected edges, are of interest, the actual scan duration required to maintain FC accuracy will be much larger than the participant-level averages shown in Figure \ref{fig:durationChange}A. This is especially true in 36P-processed data, where we observe greater heterogeneity across edges compared with FIX-processed data.

\subsection{Repeated measures data disentangles motion artifact from motion trait}
\label{sec:QCFC}

%Figure 4: Repeated Measures QC-FC

\begin{figure}
    \centering
    \includegraphics[width=\textwidth]{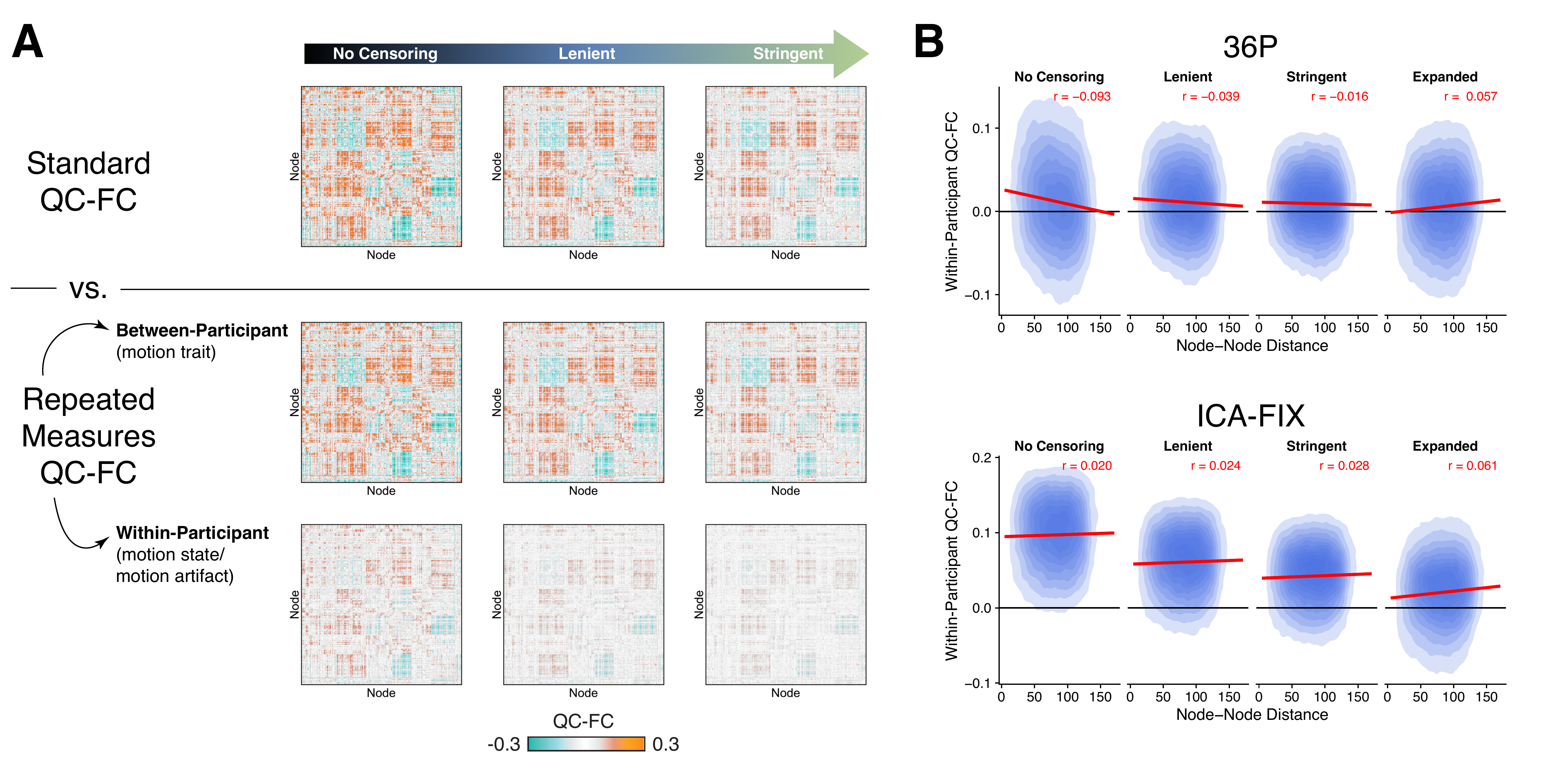}
    \caption{\small \small \textbf{Standard QC-FC is dominated by motion trait, while repeated-measures QC-FC isolates motion artifact, which stringent censoring is sufficient to mitigate.} \textbf{(A)} Repeated-measures QC-FC separates between-participant effects representing \textit{motion trait} (real between-participant differences in FC associated with motion tendency) from within-participant effects representing \textit{motion artifact} (e.g. spin history artifacts induced by head motion leading to biased FC) and possible \textit{motion state} (real within-participant fluctuations in FC associated with higher motion levels). Standard QC-FC, however, is unable to distinguish between between-participant and within-participant effects. As a result, it erroneously implies the persistence of strong motion artifacts even after stringent censoring, whereas the within-participant component of repeated-measures QC-FC shows that motion artifact is effectively mitigated after stringent censoring. \textbf{(B)} Negative distance-dependence of QC-FC, with more proximal connections showing greater associations with motion, is typically considered indicative of motion artifact.  The within-participant component of repeated-measures QC-FC shows that in 36P-processed data, distance dependence of QC-FC is reduced by lenient censoring and removed stringent censoring. Expanded censoring, on the other hand, increases the magnitude of within-participant motion-FC associations and reverses the distance effect, suggesting that it is not simply removing motion artifact. In FIX-processed data, there is no negative distance-dependence of QC-FC. Censoring does reduce the magnitude of QC-FC in FIX-processed data, possibly due to the exclusion of epochs representing motion state.}
    \label{fig:QCFC}
\end{figure}

Our findings that censoring is detrimental for FC estimation led us to revisit a conventional framework used to assess motion-related bias in FC.  Much of the justification for aggressive censoring derives from a quality control measure known as QC-FC, defined as the correlation between head motion and FC across participants \citep{power2015recent, ciricBenchmarkingParticipantlevelConfound2017}. Lower QC-FC values are thought to indicate reduced levels of motion artifact, based on the premise that associations between motion and FC are attributable to motion artifact. However, standard QC-FC is typically computed from cross-sectional data, which conflates two distinct sources of association: stable between-participant differences reflecting \textit{motion trait} (i.e., individual differences in FC associated with motion tendency) \citep{vandijkInfluenceHeadMotion2012, zeng2014neurobiological} and within-participant differences reflecting \textit{motion artifact} (e.g., spin-history effects), as well as possible \textit{motion state} (i.e. intra-individual fluctuations in FC associated with varying levels of motion tendency during different sessions). Using repeated-measures data from the HCP main cohort ($N=1087$) and longitudinal modeling techniques \citep{guillaume2014fast}, we partitioned QC-FC into between-participant and within-participant components (see Methods Section \ref{methods:QCFC}).

Standard QC-FC closely resembled the between-participant component of repeated-measures QC-FC (Figure \ref{fig:QCFC}A), indicating that conventional QC-FC is dominated by motion trait. In contrast, the within-participant component, which isolates motion artifact and/or motion state, was substantially reduced after stringent, non-expanded censoring. Thus, while standard QC-FC suggests persistent motion–FC associations even after censoring, the repeated-measures framework shows that within-participant motion effects are effectively mitigated with stringent censoring.

We next examined the distance dependence of QC-FC (Figure \ref{fig:QCFC}B). Motion artifact is known to exhibit negative distance dependence, with short-range connections being inflated due to head motion. Therefore, it is common practice to examine the relationship between QC-FC and region-to-region distance for negative associations, which are consistent with motion artifact.  In 36P-processed data without censoring, the within-participant QC-FC component exhibited a negative association with node-to-node distance ($r = -0.093$), consistent with greater motion effects in short-range connections. Lenient censoring reduced this association ($r = -0.039$), and stringent censoring essentially eliminated it ($r = -0.016$). Expanded censoring, however, increased the magnitude of within-participant motion–FC associations and reversed the distance effect ($r = 0.057$). In FIX–processed data, there was no negative distance dependence even prior to censoring ($r = 0.020$). Lenient and stringent censoring reduced the overall magnitude of within-participant QC-FC associations, while expanded censoring increased them and produced a stronger positive distance effect ($r = 0.061$)

Together, these results indicate that standard QC-FC primarily reflects between-participant motion trait. When within-participant effects are isolated through repeated-measures QC-FC, stringent censoring substantially reduces motion-related FC associations in 36P-processed data. The lack of a negative QC-FC distance effect prior to censoring in FIX-processed data suggests that motion artifact is effectively mitigated with ICA-FIX without the need for censoring, in line with previous findings \citep{pruim2015evaluation}.

It is noteworthy that expanded censoring induces a positive QC-FC distance association in both 36P- and FIX-processed data (Figure \ref{fig:QCFC}B). That is, when expanded censoring is applied, there appear systematic differences in FC within an individual across lower and higher motion sessions. One possible explanation is that when motion-proximal volumes and short epochs between motion bursts are removed, this excludes time windows representing cognitive states associated with motion tendency, thus altering FC. 

\subsection{Expanded censoring leads to attenuated and less powerful BWAS}
\label{sec:BWAS}

%Figure 5: BWAS attenuation

\begin{figure}
    \centering
    \includegraphics[width=\textwidth]{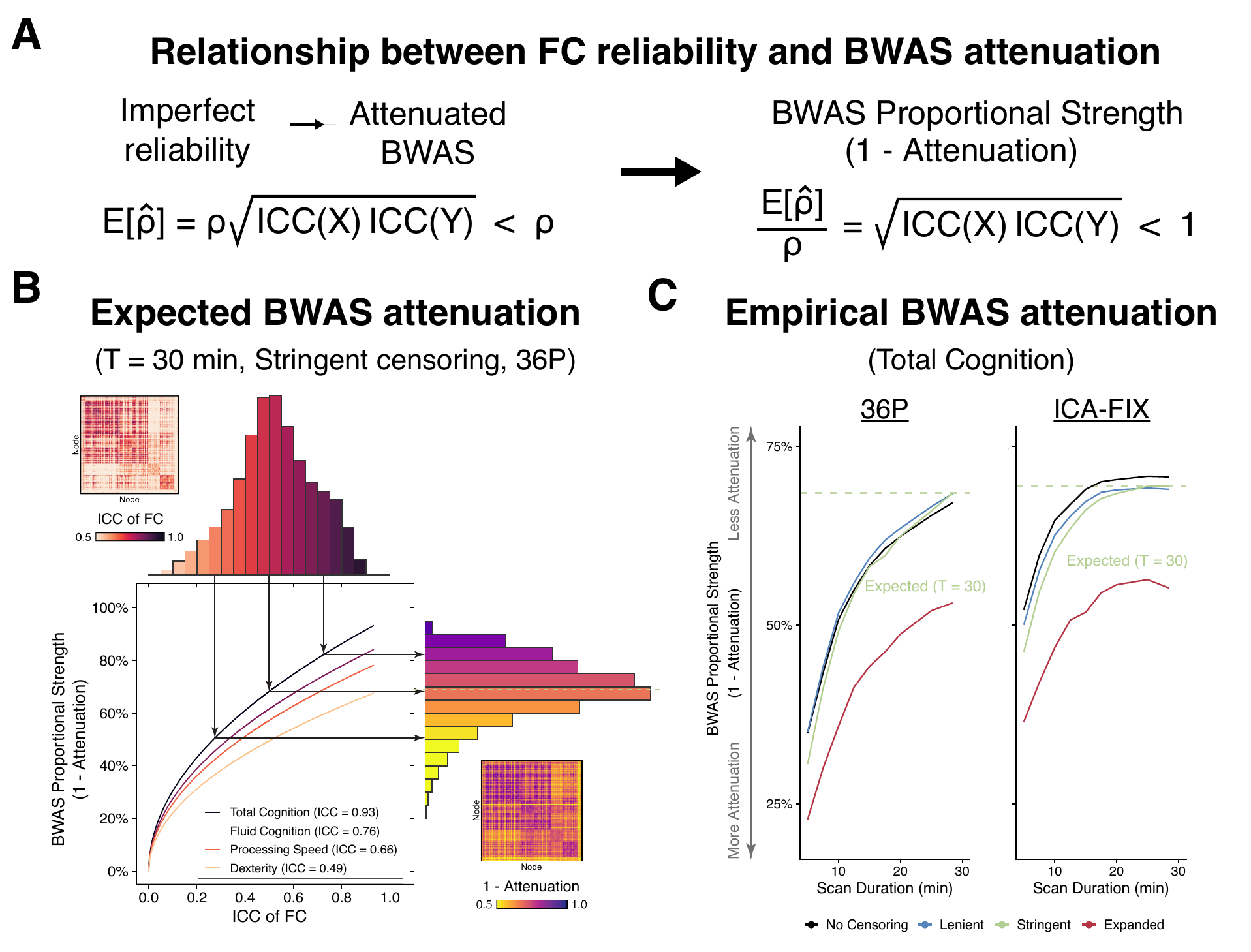}
    \caption{\small \textbf{BWAS correlations are severely attenuated due to poor FC reliability and are worsened by expanded censoring and short scan duration.} \textbf{(A)} Mathematically, BWAS correlations between imperfectly reliable brain ($X$) and behavioral ($Y$) measures are downwardly biased, and the amount of bias is determined by the reliability of both variables. We define \textit{BWAS proportional strength} as the multiplicative downward bias in BWAS associations, ranging from $0$ (total attenuation) to $1$ (no attenuation). \textbf{(B)} Illustration of the relationship between reliability of FC and behavior and BWAS attenuation. For FC, ICC is computed based on 30-minute, 36P-processed scans with stringent censoring applied. Several behavioral variables are shown, ranging from high reliability (total cognition) to low reliability (dexterity) (see Figure \ref{fig:ICC_demo} for ICC values of additional behavioral variables in the HCP). \textbf{(C)} Empirical BWAS attenuations between FC and total cognition by scan duration and censoring level, relative to ``ground truth'', bias-corrected BWAS correlations (Methods Section \ref{methods:BWAS:empirical}). Values correspond to the mean across all edges. Total cognition is the most reliable behavioral variable in the HCP, so these results represent a best-case scenario. BWAS attenuation is severe in 36P-processed data, with well over 50\% attenuation for scans up to 10 minutes without expanded censoring, and over 75\% attenuation for 5-minute scans with expanded censoring. Across all scan durations, expanded censoring severely worsens BWAS attenuation in both 36P- and FIX-processed data. In FIX-processed data, BWAS is less but still substantially attenuated, with approximately 50\% attenuation in 5-minute scans without expanded censoring and approximately 60\% attenuation with expanded censoring. The least BWAS attenuation, approximately 30\%, is seen with 30-minute FIX-processed scans with no censoring.}
    \label{fig:BWAS_attenuation}
\end{figure}

The detrimental effects of aggressive censoring on FC accuracy have downstream implications for brain-wide association studies (BWAS), which center around correlations between FC and behavioral measures. When either of those measures is unreliable, estimated BWAS correlations will be systematically attenuated toward zero and more variable (Methods Section \ref{methods:BWAS}).  Here we examine the relationship of censoring to BWAS attenuation and variance. 

The magnitude of BWAS attenuation depends on the reliability of both FC and the behavioral variable (Figure \ref{fig:BWAS_attenuation}A). We quantify this effect by defining BWAS \textit{proportional strength} as the expected sample correlation relative to the true correlation. It is equivalent to the classical attenuation factor of a sample correlation between two variables $X$ and $Y$, given by $\sqrt{\text{ICC}(X)\text{ICC}(Y)}$, which ranges from 0 (complete attenuation) to 1 (no attenuation). Note that sample size has no effect on BWAS attenuation, as this downward multiplicative bias depends only on the reliability of FC and behavior, not the number of participants.  

Figure \ref{fig:BWAS_attenuation}B  illustrates the expected BWAS proportional strength based on observed FC reliability with 30-minute scans and stringent censoring in 36P-processed data, for a range of behavioral measures. Behavioral reliability varies widely in the HCP (Figure \ref{fig:ICC_demo}). Only total cognition exhibits excellent reliability (ICC = 0.935), whereas most other measures show moderate or good reliability \citep{koo2016guideline}. Even with a highly reliable behavior (total cognition) and long scan duration, imperfect FC reliability results in substantial attenuation. Lower behavioral reliability further reduces BWAS proportional strength.

We next evaluated empirical BWAS attenuation using total cognition and bias-corrected “ground truth” BWAS correlations (Figure \ref{fig:BWAS_attenuation}C). In 36P-processed data, BWAS attenuation was substantial across scan durations. Without expanded censoring, scans up to 10 minutes exhibited well over 50\% attenuation. With expanded censoring, attenuation exceeded 75\% for 5-minute scans and remained severe across all durations. In FIX–processed data, attenuation was less extreme but remained substantial. Five-minute scans without expanded censoring showed approximately 50\% attenuation, increasing to approximately 60\% with expanded censoring. The least attenuation (approximately 30\%) was observed for 30-minute ICA-FIX scans without censoring. 

Poor reliability has another negative implication for BWAS correlations: in addition to attenuating their magnitudes, poor reliability also results in higher sampling variance (Methods Section \ref{methods:BWAS}). BWAS sampling variance primarily depends on sample size (more participants, lower variance) and the magnitude of the \textit{expected} correlation (smaller magnitude, higher variance) (Figure \ref{fig:BWAS_var_math}). Thus, imperfect reliability increases the sampling variance of BWAS correlations indirectly by attenuating their average magnitude.  In a theoretical example assuming a true correlation of 0.5 and behavioral reliability corresponding to total cognition (ICC $\approx$ 0.93), the sample size required to achieve a given level of BWAS variance differed substantially depending on FC reliability (Figure \ref{fig:BWAS_var_math}). When FC reliability was high (95th quantile across edges), approximately 1,100 participants were required to achieve the same BWAS variance as approximately 1,500 participants when FC reliability was low (5th quantile across edges). Thus, differences in FC reliability can translate into differences of hundreds of participants required to obtain equivalently stable BWAS estimates.

Together, these results show that reduced FC reliability due to short scan duration, suboptimal processing, and/or aggressive censoring can undermine brain-wide association studies by attenuating the magnitude of BWAS correlations and increasing their error variance. This ultimately inflates sample size requirements, since more participants are needed to offset the higher variance due to suboptimal FC reliability. However, increasing sample size cannot mitigate BWAS attenuation, which will be persistent unless steps are taken to maximize the reliability of both FC and behavior.

\section{Discussion}

In this study, we examined the impacts of head motion-based volume censoring of fMRI data on downstream analyses and data quality. Specifically, we considered the effect of censoring on the accuracy of individual-level functional connectivity (FC) estimates, levels of noise and motion-induced bias, and the strength and reliability of brain-behavior associations in population-wide brain-wise association studies (BWAS). Across these lines of inquiry, we consistently observed more aggressive censoring practices to be detrimental. Censoring degraded FC accuracy due to excessive data loss and undermined BWAS through systematic attenuation of brain-behavior correlations. We also found that conventional ways of evaluating the effects of motion, which rely on associations between motion and FC, are largely driven by motion trait and tend to exaggerate levels of motion artifact.  While mitigating the impacts of motion remains essential for fMRI studies, our findings demonstrate the importance of considering the balance between noise removal and data retention to ultimately optimize the fidelity of downstream analyses. In many cases, aggressive censoring imposes a high cost without commensurate measurable benefits.

\subsection{Motion state, and the challenge of distinguishing artifact from cognition}

Our findings suggest that aggressive motion censoring removes more than just {motion artifact}, i.e., head motion-induced contamination of the BOLD signal due to disruption of proton spin history.  In particular, the alterations to within-participant motion-FC associations observed with expanded censoring suggest the selective removal of certain FC patterns occurring in temporal proximity to motion, i.e. \textit{motion state}. This distinction raises an important conceptual issue. If motion-proximal FC patterns reflect true neurobiological states linked to arousal, cognitive engagement, or behavioral traits, then excluding or under-sampling them may remove meaningful signal. On the other hand, if the goal is to isolate FC patterns unrelated to motion tendency, one might argue that removing motion-related FC states is desirable. Yet even in that case, expanded censoring is neither the only, nor the most targeted, strategy for achieving that objective. 

Dynamic FC approaches may offer a more principled alternative.  By explicitly modeling time-varying connectivity and identifying recurrent FC states, dynamic FC analyses allow comparisons within matched states across individuals. This framework avoids discarding data while recognizing the benefit of isolating motion state.  Further, it separates state occupancy (e.g., dwell time in motion state versus other cognitive states) from state-specific connectivity patterns. Dynamic FC approaches may therefore provide a more nuanced solution to addressing motion state than aggressive censoring. 

An important implication of our findings is the need to move beyond simple motion-FC associations toward more causal frameworks to evaluate motion artifact.  In this study, we advanced in this direction through a simple yet powerful framework for QC-FC based on repeated measures data, allowing us to distinguish motion trait from motion artifact or state. Indeed, we find motion trait to be the primary driver of conventional QC-FC based on cross-sectional data, underscoring the importance of distinguishing it from less desirable motion-FC associations. However, repeated measures analyses cannot distinguish true motion artifact from genuine FC patterns associated with time-varying motion tendency, i.e. motion state. More rigorous causal and experimental frameworks are therefore needed to isolate the causal effects of head motion on the BOLD signal and on functional connectivity.  This will be essential for refining motion mitigation strategies and ensuring that denoising methods effectively target motion artifact without inadvertently removing true neural signal.

\subsection{Implications for BWAS, and the limits of sample size}

Our findings have important implications for brain-wide association studies.  Prior work has centered around the \textit{reliability} of brain-behavior correlations \citep{Marek2022} and the need for very large sample sizes consisting of thousands of participants. By contrast, we consider both reliability and \textit{validity}, distinguishing between the variance and bias of estimated correlations, and the examining drivers of each. The variance, sample-to-sample fluctuations around the expected value, is primarily reduced by increasing sample size (though it is also influenced by FC reliability, as discussed below). Downward bias, on the other hand, arises due to imperfect reliability of FC and behavior. Increasing sample size does not reduce this bias; rather, it stabilizes estimates around an attenuated value. As a result, even very large studies may produce highly reliable yet systematically underestimated brain-behavior associations.  Achieving BWAS correlations that are not only reliable but also valid therefore requires attention to both data quantity and data quality. 

We find that the least attenuated BWAS correlations result from increasing scan duration, adopting advanced denoising techniques such as ICA-FIX, and avoiding unnecessary data loss from aggressive censoring. In data denoised with ICA-FIX, BWAS attenuation essentially plateaued after 15 minutes of scan duration, whereas data denoised with 36P confound regression was highly sensitive to scan duration beyond 15 minutes. Because large population studies can rarely acquire long scans, adopting advanced denoising techniques may allow for moderately short acquisitions (though ideally not below 15 minutes) without compromising BWAS.  Given a fixed data collection budget, reducing scan duration requirements while maintaining FC reliability may allow for increasing sample size, thereby enhancing BWAS reliability without compromising validity. 

Poor FC reliability has implications beyond attenuation in BWAS: it also inflates the variance of BWAS correlations. That is because variance is determined by two factors: sample size (larger samples reduce variance) and the expected value of the sample correlation (smaller expected value increases variance). Any factor that attenuates BWAS correlations also inflates their variance, thus increasing the sample size required to achieve reliable BWAS. Because scan duration is a major determinant of FC reliability, overly short acquisitions in large population studies, in conjunction with suboptimal processing choices, imposes substantial hidden costs by increasing sample size requirements. 

In principle, BWAS attenuation can be corrected when the reliability of FC and behavior are known \citep{spearman1910correlation, nikolaidis2022suboptimal}. Such bias correction rescales estimated correlations to account for imperfect measurement and can serve to improve comparability of brain-behavior associations across edges with differential reliability. However, correction depends on accurate estimates of reliability and can increase sampling variance when those estimates are noisy. Therefore, improving FC reliability through judicious processing choices and sufficient scan duration remains preferable to post-hoc correction. It is also critical to improve the reliability of behavioral measures, which contribute equally to BWAS attenuation and may be more feasible to optimize than FC given practical constraints on scan duration. 

Finally, censoring can also affect BWAS through participant exclusion. Aggressive censoring often leaves insufficient data remaining for some individuals, leading to their exclusion from downstream analyses. Discarding data from participants with poor-quality FC can be beneficial, as it improves FC reliability within the sample and thereby reduces BWAS attenuation. However, exclusions simultaneously reduce sample size, thus increasing the variance of brain-behavior correlations and diminishing statistical power. Moreover, systematic exclusions create a less representative sample and compromise generalizability. While some exclusions are unavoidable due to severe noise contamination, our findings suggest that more judicious approaches to denoising can effectively mitigate motion artifact while avoiding the excessive data loss and participant exclusion rates associated with aggressive censoring.

\subsection{Implications for clinical fMRI}

Our findings also have implications for clinical and translational applications of fMRI, which depend on individual-level precision.  In clinical contexts, such as presurgical mapping of essential brain functions or MRI-guided brain stimulation \citep{kong2025network, kong2026efficacy}, patients who have difficulty controlling movement may be disproportionately affected by motion censoring.  Aggressive censoring substantially increases the necessary scan duration to achieve high levels of reliability from fMRI-based measures. This may necessitate longer or repeated sessions for some patients. Adopting more lenient censoring strategies, or using advanced denoising techniques in lieu of censoring, can preserve data while effectively mitigating motion artifact, thereby reducing the burden and cost to patients.

\subsection{The value of image visualization}

Functional neuroimaging has evolved from its early focus on task-based activation mapping to include FC and broader network neuroscience approaches. FC matrices and network metrics offer complementary, systems-level views that have greatly advanced our understanding of brain function and organization. However, exclusive reliance on FC matrices, network metrics, and other high-level abstractions also risks moving researchers further away from the underlying image data, weakening awareness of its quality and trustworthiness. FC matrices and network diagrams do not readily allow visual assessment of noise levels or data quality. In contrast, activation maps or seed connectivity maps provide intuitive indicators of these factors, such as spatial smoothness and coherence. Notably, it is rare for researchers to utilize first-level task activation maps in downstream analyses due to perceived high noise levels; yet this is common practice in analyses based on FC matrices, potentially reflecting a lack of awareness of their trustworthiness, or the importance thereof. Thus, complementing FC analyses with image-based representations, such as participant-level seed connectivity maps and population-level summaries (e.g., variance, signal-to-noise ratio), can provide important visual feedback and help elucidate how processing choices influence the balance between signal and noise.

\subsection{Limitations and future directions}

Several limitations of our study should be noted. First, since our primarily focus here is motion censoring, we considered just two denoising strategies: confound regression and ICA-FIX. Other popular denoising methods, such ICA-AROMA or aCompCor, are also worth consideration, especially because ICA-FIX may not be feasible in all settings. The FIX classifier requires manual training, as well as sufficient data to ensure accurate component classification. These requirements may pose challenges for smaller datasets, certain clinical populations, or multi-site studies with heterogeneous acquisition protocols. Furthermore, we did not consider purely data-driven censoring approaches, which may offer improved specificity and discard less data, especially when computed after denoising. We also did not consider despiking, which addresses burst noise without discarding entire volumes. Future work is needed to comprehensively evaluate the effect of these various denoising techniques on individual precision and BWAS. 

Second, we did not evaluate the impact of band-pass temporal filtering, which is common in resting-state fMRI studies. Instead, we used only high-pass filtering, as foregoing a low-pass filter has been shown to improve signal-to-noise separability and reliability in resting-state analyses \citep{shirerOptimizationRsfMRIPreprocessing2015}. Band-pass filtering can result in considerable loss of temporal degrees of freedom---approximately 60\% for 2s TR and 80\% for 1s TR \citep{reynolds2024processing}---largely due to removal of higher-frequency fluctuations. This is an important but largely overlooked consequence of band-pass filtering. Given the importance of data retention for maximizing FC accuracy and minimizing BWAS attenuation, additional data loss associated with low-pass filtering may have meaningful downstream effects. Future work should examine how band-pass versus high-pass filtering interacts with motion mitigation strategies to influence individual-level FC accuracy and BWAS reliability.

\subsection{Recommendations}

Our results demonstrate advantages of ICA-FIX over 36P confound regression for mitigating noise, with little added benefit of motion censoring when ICA-FIX is applied. Although ICA-FIX may not be feasible in all settings, as discussed above, these findings underscore the value of advanced denoising techniques that effectively mitigate motion artifact without the need for censoring. Other ICA-based approaches \citep{pruim2015evaluation} and multi-echo denoising techniques \citep{reddy2024denoising} have similarly shown strong denoising efficacy without censoring. In addition, robust statistical approaches, such as rank-based correlations, may also warrant consideration as an alternative to removal of volumes.

\section{Methods}
\label{sec:methods}

\subsection{Datasets}

We used resting-state functional MRI (rs-fMRI) data from the Human Connectome Project (HCP) \citep{van2013wu}.  At each visit, two runs of 1200 volumes lasting approximately $15$ minutes were collected: one with left-to-right (LR) phase encoding and the other with right-to-left (RL) phase encoding.  
For ground truth functional connectivity (FC) analysis described in Sections \ref{sec:FC_validity}-\ref{sec:budget_inflation}, we utilized data from the HCP Retest dataset, which includes data from 45 participants from the main HCP who underwent the protocol a second time.  Participants in the HCP Retest sample therefore underwent 8 resting-state fMRI (rsfMRI) runs in total across four visits.  Three retest participants were omitted due to missing or incomplete rsfMRI data (627549, 341834 and 143325), leading to a sample size of $n=42$. Four additional participants were excluded because they did not have enough data to define ground truth FC (see next section), so the final sample size was 38. 
% omitted HCP retest participants:
% -  627549, # does not have retest zip files
% -  341834, # does not have retest FIX files in the zip (REST1 RL)
% -  143325  # retest MPP CIFTI for REST2_RL is truncated to 939 timepoints

For QC-FC (Section \ref{sec:QCFC}) and the brain-wide association study (BWAS) analysis (Section \ref{sec:BWAS}), we utilized data from the 1200-participant release of the young adult HCP, which includes functional and structural MRI data on 1095 participants.  Eighteen participants who were missing all four rs-fMRI runs were excluded, resulting in a sample size of $n = 1087$.

\subsection{Data processing and censoring}
\label{methods:scrub_and_denoise}

For each fMRI run, we first dropped the first 15 volumes (10 seconds) to allow the scanner to reach equilibrium. We then parcellated the data by calculating the average timeseries within each of 400 cortical parcels \citep{schaeferLocalGlobalParcellationHuman2018} and within each of 19 Freesurfer parcels representing subcortical and cerebellar regions \citep{fischl2012freesurfer} (Figure \ref{fig:region_labels}). Prior to computing FC, we denoised each fMRI run using one of two methods: minimal preprocessing \citep{glasserMinimalPreprocessingPipelines2013} 
followed by nuisance regression with 36 regressors (36P) \citep{satterthwaiteImprovedFrameworkConfound2013}, or ICA-FIX as implemented in the HCP data release \citep{griffanti2014ica}.  Briefly, 36P includes 6 realignment parameters, 3 mean signals (white matter, cerebral spinal fluid, and global signal within grayordinates), along with the squares, temporal differences, and squared temporal differences of those 9 signals.  Nuisance regression with 36P was applied after parcellation for computational efficiency, which is mathematically equivalent to performing nuisance regression voxel- or vertex-wise first, then parcellating.
To examine the effect of scan duration, we created scans with shorter duration $T$ varying from $5$ to $25$ minutes.  For a given visit and scan duration $T$, we took the first $T/2$ volumes from the LR and RL acquisitions, respectively. 

For censoring, we used a lagged and filtered version of framewise displacement (FD) appropriate for multiband data, as described in \cite{pham2023less}.  FD thresholds ranging from 0.5mm (``FD5'', lenient) to 0.2mm (``FD2'', stringent) were considered.
In addition, we computed DVARS per \cite{afyouni2018insight}, which provides a statistically principled, adaptive method for identifying volumes with high DVARS, thus avoiding the need for a user-chosen threshold. To increase sensitivity to noise, DVARS was computed on the minimally-preprocessed CIFTI data.  Since the DVARS thresholding technique is adaptive, we computed the DVARS threshold separately for each scan duration of each run to avoid data leakage.

For ``expanded'' censoring, for each volume flagged due to high FD or high DVARS, we censored one preceding volume and two subsequent volumes, as well as any remaining segments containing fewer than 5 volumes.  We only consider expanded censoring in combination with the stringent FD threshold of 0.2mm, since the practice of expansion typically accompanies stringent FD thresholds. Figure \ref{fig:FD_Comparison} illustrates expanded, stringent censoring compared with non-expanded stringent censoring.

For 36P-processed data, nuisance regression, high-pass temporal filtering, and censoring were performed in a simultaneous regression framework as recommended by \cite{lindquistModularPreprocessingPipelines2019}. We did not perform band-pass filtering, since removal of high-frequency components leads to excessive loss of degrees of freedom \citep{reynolds2024processing}, and has been shown to worsen reliability and signal-to-noise separation in resting-state fMRI analyses \citep{shirerOptimizationRsfMRIPreprocessing2015}. For ICA-FIX processed data, the previously denoised data were high-pass filtered and censored using simultaneous regression. In order to incorporate high-pass filtering in the simultaneous regression, we included a set of discrete cosine basis (DCT) functions to achieve a high-pass filter of 0.01 Hz. The DCT basis set for a particular frequency cutoff, scan duration, and TR were generated using the \texttt{dct\_bases} function in the \texttt{fMRItools} R package.  Therefore, for each run and scan duration, the single, simultaneous nuisance regression thus included the intercept, nuisance parameters, DCT basis functions, and spike regressors \citep{satterthwaiteImprovedFrameworkConfound2013} for censored volumes.  To avoid data leakage, we performed this regression separately for each scan duration for each run.  Following nuisance regression, volumes to be censored were removed prior to FC calculation.

\subsection{FC calculation}
\label{methods:FC}

For each run and scan duration, pairwise FC between the 419 parcels was computed as Pearson correlation, following denoising and removal of censored volumes, as described above. FC values were Fisher transformed prior to further analysis. For each visit and scan duration $T$, FC was averaged over LR and RL runs (using the first $T/2$ minutes of each run). 

\subsubsection{Ground truth FC}
\label{methods:ground_truth}

For each HCP Retest participant included in our analyses, eight 15 minute rest fMRI (rfMRI) runs were acquired, for a total of nearly 2 hours per participant. To establish the ground truth FC for each participant, we used 6 runs (1.5 hours), to which we applied stringent censoring. The 6 ground truth runs included both retest visits (4 runs) and one of the two main HCP visits (2 runs). The ground truth FC was computed as the weighted average over all 6 runs, weighted by the number of uncensored volumes per run. The remaining main HCP visit (30 minutes) was held out for FC estimation based on typical scan durations (5 to 30 minutes), with or without censoring. This entire analysis was repeated by switching which main HCP visit was used in the ground truth and which was held out.  After computing FC error for each split (see below), we computed the weighted average over both splits, where each split was weighted by the number of uncensored volumes in the held out data, for a given scan duration. 

After stringent censoring, three participants had less than 50 minutes remaining of the original 1.5 hours in one or both partitions. These participants had the IDs 103818, 151526, and 175439, and were excluded from all analyses, as insufficient data was left to determine the ground truth with a high degree of accuracy. For the remaining 39 participants, we computed the ground truth FC as the weighted average of the Fisher-transformed FC estimates across the 6 runs.  

\subsubsection{FC Error}

The accuracy of participant-level FC estimates was assessed based on the error relative to the participant's ground truth FC. We first excluded any held-out sessions where less than 150 volumes remained after expanded censoring for the 5-minute scan duration. This resulted in exclusion of 1 session from each of 4 participants but did not lead to removal of any participants. For each participant, the absolute error (absolute difference between the estimate and the ground truth) was averaged over both ground truth partitions, weighted by the number of uncensored volumes contributing to the partition-specific estimates. The mean absolute error was then squared to produce squared error, which was then averaged over participants and/or edges to produce mean squared error (MSE). Figures show the square root of the MSE (rMSE), which is on the same scale as FC. For tests of significant differences in FC error across censoring levels, we computed the rMSE across edges for each participant and applied robust paired Wilcoxon tests, thus avoiding the influence of outlier participants. Note that these tests consider participants, but not edges, independent, and make no distributional assumptions. 

Note that as a result of the exclusion of nearly 10\% of participants and the exclusion of some held-out sessions with insufficient data remaining, the negative effects of aggressive censoring on FC accuracy in Sections \ref{sec:FC_validity}-\ref{sec:budget_inflation} may be somewhat understated, since we find that over-censoring is more detrimental for high movers (Section \ref{sec:worse_for_high_movers}).

\subsection{Noise Variance}
\label{methods:baseline_var}

In Section \ref{sec:worse_for_high_movers}, we examine the effect of censoring on two competing factors driving participant-level FC estimation error: noise levels in the data and (effective) scan duration.  In this section, we explain how we estimate both of those quantities.

\subsubsection{Noise Variance Model}

To determine noise levels, consider a simple measurement error model for FC.  For participant $i$ and edge $j$, letting $w_{ij}$ be the true FC and $x_{ij}^{(T)}$ be the estimated FC based on $T$ volumes,
$$
x_{ij}^{(T)} = w_{ij} + e_{ij}^{(T)},
$$
where $e_{ij}^{(T)}$ has mean zero and variance $Var(e_{ij}^{(T)})$, the estimation error variance. For an unbiased estimator, the error variance is equal to the mean squared error (MSE) of $x_{ij}^{(T)}$ relative to the truth $w_{ij}$. This variance is influenced by multiple factors, including scan duration, autocorrelation, and other variables, but can be reasonably assumed inversely related to the (effective) scan duration $T_{\mathrm{eff}}$:
$$
\Var(e_{ij}^{(T)}) = \sigma^2_{ij}/T_{\mathrm{eff}},
$$
where $\sigma^2_{ij}$ is the \textit{noise variance} in the fMRI data.  Effective censoring reduces noise (good for estimation error) while decreasing scan duration (bad for estimation error). These two competing forces may ultimately result in lower or higher error variance $\Var(e_{ij}^{(T)})$.  We empirically assess the impact of censoring on the noise variance for acquired scan duration $T=10$ minutes. For censoring method $m$, noise variance $\sigma^{2}_{ijm}$ is estimated as
$$
\sigma^{2}_{ijm} = T_{im} \times \Var\big(e_{ijm}^{(T)}\big),
$$
where $T_{im}$ is the effective scan duration (see below) for participant $i$ after censoring with method $m$, and $\Var(e_{ijm}^{(T)})$ is determined empirically using the HCP Retest dataset, as the MSE between estimated and ground truth FC values (see Section \ref{methods:FC}).

\subsubsection{Effective Scan Duration}
\label{methods:ESS}

Censoring affects the temporal correlation structure of the data and hence the effective scan duration. For instance, removing every other time point does not discard half of the information in autocorrelated data, since the retained time points are correlated with the discarded time points. Therefore, instead of using the nominal scan duration after censoring to determine the noise variance, we estimate the \textit{effective} scan duration as follows. 

First, we compute the empirical autocorrelation function (ACF) of the BOLD timeseries of each parcel using the \texttt{acf} function in R, then average across parcels. ACFs for lags over 100 are set to zero. Outlying volumes may influence the ACF, but typically have a minor effect since all volumes contribute to the ACF at each lag. Indeed, a robust version of the ACF using the \texttt{robacf} function in R produced nearly identical results (not shown). However, to further mitigate the effect of any outliers, we take the median ACF over participants and runs. 

For a stationary autocorrelation process, the $T\times T$ temporal correlation matrix $\boldsymbol\Sigma$ is a Toeplitz matrix, with the $\ell$th off-diagonal containing the lag-$\ell$ autocorrelation. Thus, the ACF estimated as above determines $\boldsymbol\Sigma$ \textit{prior to} censoring. To determine the temporal correlation matrix \textit{after} censoring, the rows and columns corresponding to the censored volumes are simply removed from $\boldsymbol\Sigma$ to produce a matrix representing the correlation among the remaining volumes. 

Effective sample size (ESS) is often defined as the number of independent samples needed to achieve the same estimation efficiency (inverse of the sampling variance) as that based on the observed, potentially correlated, samples. While ESS is often based on the efficiency of the sample mean, here we are interested in estimating not the mean, but the correlation between time series. \cite{afyouni2019xDF} derived the sampling distribution of FC in the presence of autocorrelation in the general case. It can be shown that in our special case of homogeneous temporal autocorrelation and zero mean, the sampling variance of an FC estimate $\hat\rho$ with true value $\rho$ is given by 
$$
\Var(\hat\rho_\Sigma) = {(1-\rho^2)^2}\text{Tr}(\boldsymbol\Sigma^2)/T^2,
$$ 
where $\text{Tr}(\mathbf{A})$ denotes the trace of the matrix $\mathbf{A}$. By comparison, the sampling variance without autocorrelation (i.e., $\boldsymbol\Sigma = \mathbf{I}_T$) is $\Var(\hat\rho_I)={(1-\rho^2)^2}/T$. Note that $\Var(\hat\rho_\Sigma)\geq \Var(\hat\rho_I)$, since $\text{Tr}(\boldsymbol\Sigma^2) \geq T$ and only equals $T$ when $\boldsymbol\Sigma = \mathbf{I}_T$. The effective scan duration, defined as the nominal sample size $T$ multiplied by the relative efficiency (the ratio of the variances) in the autocorrelated case, is thus given by
$$
T_{\mathrm{eff}} 
= T \times \frac{\Var(\hat\rho_I)}{\Var(\hat\rho_\Sigma)}
= \frac{T^2}{\text{Tr}(\boldsymbol\Sigma^2)}
$$ 

To illustrate the extreme cases: if there is no autocorrelation, i.e. $\boldsymbol\Sigma=\mathbf{I}_T$, then $\text{Tr}(\boldsymbol\Sigma^2)= \text{Tr}(\mathbf{I}_T) = T$, so $T_{\mathrm{eff}} =T$; at the other extreme, if all time points $1$ to $T$ are perfectly correlated, then $\boldsymbol\Sigma$ is a matrix of $1$s, so $\text{Tr}(\boldsymbol\Sigma^2) = T^2$, so $T_{\mathrm{eff}} = 1$.  If there is autocorrelation to a lesser degree, $T_{\mathrm{eff}}$ will be less than $T$. 

Using this definition of effective scan duration, we compute it for each participant, run and censoring method, for nominal scan duration $T = 10$ minutes.

\subsection{Required Scan Duration}
\label{methods:required_scan_duration}
Using the same data as in the analyses shown in Figure \ref{fig:FC_accuracy}, \textbf{Figure \ref{fig:durChange_illustration}} illustrates how we compute the required scan duration to maintain FC accurary, relative to lenient censoring with a scan duration of $T=17.5$ minutes. This scan duration was chosen because it is the midpoint of the range of $5$ minutes to nearly $30$ minutes of observed scan durations, providing an equal range of possible increases or decreases in required scan duration.

\subsection{Repeated Measures QC-FC}
\label{methods:QCFC}

QC-FC refers to the association between head motion and functional connectivity (FC).  A limitation of QC-FC computed from cross-sectional data is that it does not distinguish \textit{motion trait}, i.e. patterns of FC associated with a participant's propensity to move, from motion artifact \citep{vandijkInfluenceHeadMotion2012, zeng2014neurobiological}.  To account for motion trait, it is necessary to separate across-participant associations between motion and FC from within-participant associations more likely reflective of motion artifact. We propose a new version of QC-FC based on repeated measures data, which identifies both between-participant motion-FC associations and within-participant motion-FC associations. To accomplish this, we adopt tools from longitudinal analysis designed estimate the cross-sectional (between-participant) and longitudinal (within-participant) associations using a linear regression framework. 

First, note that standard QC-FC, given by the correlation between motion and FC, can be equivalently computed via linear regression: it is the slope coefficient in an intercept-free simple linear regression model relating FC, $y_i$, to FD, $x_i$, participants $i=1,\dots,n$, where both the $x_i$ and the $y_i$ have been centered and scaled to unit variance. With access to repeated measures for each participant, that simple linear regression model can be extended to separately estimate within-participant and between-participant effects. This is accomplished by creating two FD regressors, one for the per-subject FD average, and another for the subject-centered FD \citep{guillaume2014fast}. Let $x_{ij}$ represent the motion level (mean FD) for participant $i$ at session $j$, and let $y_{ijk}$ represent the corresponding FC at edge $k$.  Let $\bar{x}_i$ be the mean of $x_{ij}$ across sessions for participant $i$, and let $\bar{\bar{x}}$ be the overall mean across all sessions and participants. Define regressors 
$$
x_{ij}^{\text{within}}=x_{ij} - \bar{x}_i
\quad\text{and}\quad 
x_{ij}^{\text{between}} = \bar{x}_i - \bar{\bar{x}},
$$
and scale both to have unit variance.  Then the coefficients in the intercept-free linear regression model relating $y_{ijk}$ to $x_{ij}^{\text{within}}$ and $x_{i}^{\text{between}}$ represent the \textit{partial} correlations between FC and FD within participants and between participants, respectively. 

Using main HCP ($n = 1095$), which includes four runs per participant, we used this approach to estimate the within- and between-participant QC-FC, along with standard QC-FC, for each edge and for each censoring level. Note that participants can be included even if some runs are missing, since the regression approach allows for missing data. We excluded any runs with less than $1$ minute of scan time remaining after expanded censoring, which resulted in the exclusion of 8 participants. Of the remaining $n=1087$ participants, 52 are missing a single run, 86 are missing two runs, and 11 are missing three runs. For all participants, the participant-specific mean $\bar{x}_i$ was computed using the available sessions.  The overall mean $\bar{\bar{x}}$ was computed as the mean across the $\bar{x}_i$ for $i=1,\dots, n$. 

As a validation of repeated measures QC-FC, we also computed the association between within-participant changes in FD and FC directly.  For each participant, we identified their highest-motion and lowest-motion run. We then computed the change in FD and the change in FC between those two runs. Since this analysis requires at least two runs per participant, participants with only a single rs-fMRI run were excluded. We then computed the Pearson correlation between the within-participant change in FC and the corresponding change in FD across the two runs. This more ``direct'' approach gives consistent results as the within-participant component of repeated measures QC-FC (Figure \ref{fig:QCFC_36P_validation}).

\subsection{BWAS Attenuation and Variance}
\label{methods:BWAS}

In brain-wide association studies (BWAS), we are interested in the association between brain measures and behavioral variables.  Let $y^\text{true}_i$ be the true behavioral measure for participant $i$, and let $x^\text{true}_{ij}$ be the true FC of participant $i$ at edge $j$. The true BWAS correlation is given by
$$
\rho_j = \Cor(x^\text{true}_{ij}, y^\text{true}_i).
$$

Using the same measurement error notation as in Section \ref{methods:baseline_var}, let $w_{ij}$ be the true FC for participant $i$ at edge $j$, let $x_{ij}^{(T)}$ be the estimated FC based on a scan duration of $T$, and let $e_{ij}^{(T)}$ be the measurement error. Let $y_i$ be the measured value of the behavioral variable.  Using the estimated FC and behavior in place of their true values, the estimated BWAS correlation is 
$$
\hat\rho_{j}^{(T)} = \Cor(x_{ij}^{(T)}, y_i).
$$

\subsubsection{BWAS Attenuation due to Imperfect Reliability}

In real settings, FC and behavior are not perfectly reliable. In that case, the estimated BWAS correlation, $\hat\rho^{(T)}_{j}$, will be biased downwards or \textit{attenuated} \citep{spearman1910correlation, gell2023burden}. Specifically, its expected value is 
$$
\mathbb{E}[\hat\rho_{j}^{(T)}] = \rho_j \sqrt{R(X_{j}^{(T)})R(Y)} < \rho_j,
$$ 
where $R(X_{j}^{(T)})$ is the intra-class correlation coefficient (ICC) of FC at edge $j$, and $R(Y)$ is the ICC of the behavioral measure. Therefore, on average, $\hat\rho_{j}^{(T)}$ will be attenuated relative to the true value $\rho_j$. Define \textit{BWAS proportional strength} as
$$
\frac{\mathbb{E}[\hat\rho_{j}^{(T)}]}{\rho_j} = \sqrt{R(X_{j}^{(T)})R(Y)},
$$
which ranges from 0 (total attenuation) to 1 (no attenuation). Note that increasing the number of participants, $N$, has no effect on BWAS attenuation, which depends only on participant-level measurement reliability.
Censoring can affect BWAS attenuation by altering the ICC of FC, which is given by
$$
R(x_{j}^{(T)}) = 
{\frac{\Var(w_{ij})}{\Var(w_{ij}) + \Var(e_{ij}^{(T)})}} ,
$$
where $\Var(w_{ij})$ is the between-participant or ``signal'' variance of the true FC at edge $j$, and $\Var(e_{ij}^{(T)})$ is the within-participant or ``error'' variance of FC estimates around the truth. 
FC error is affected by censoring, as examined empirically and theoretically above. While theoretically censoring may either reduce or increase FC error variance, as explained in Section \ref{methods:baseline_var}, empirically we find that censoring tends to increase FC error variance. Hence, censoring would tend to result in lower FC ICC and thus worse BWAS attenuation. However, censoring may also have other effects that could influence BWAS attenuation: unintentional selective sampling of certain FC states through censoring may inadvertently affect the between-participant variance $\Var(w_{ij})$; also, censoring often results in exclusion of participants due to insufficient scan time remaining, which could also affect between-participant variance.  Below, we empirically investigate the effect of censoring on BWAS attenuation for different scan durations, processing methods, and censoring levels.

\subsubsection{BWAS Variance}

The error variance of $\hat\rho_{j}^{(T)}$ is given by
$$
\Var(\hat\rho_{j}^{(T)}) 
= \frac{1}{N}\big(1 - \rho_j^2 R(x_{j}^{(T)})R(Y)\big)^2
= \frac{1}{N}\big(1 - \mathbb{E}[\hat\rho_{j}^{(T)}]^2\big)^2.
$$
Two factors therefore drive BWAS variance: sample size $N$ (increasing sample size reduces variance) and the \textit{expected} BWAS correlation, $\mathbb{E}[\hat\rho_{j}^{(T)}]$ (smaller magnitude increases variance). Therefore, the presence of BWAS attenuation also increases BWAS variance, for a given sample size. Thus, if censoring worsens BWAS attenuation by worsening FC reliability, it will likewise increase BWAS variance. Censoring can also affect BWAS variance due to participant exclusion, which reduces sample size. Thus, censoring is expected to increase BWAS variance through worse FC reliability and reduced sample size.

\subsubsection{Empirical Analysis}
\label{methods:BWAS:empirical}

To empirically evaluate the effect of censoring, processing and scan duration on BWAS attenuation, we first established ground-truth BWAS using a large sample and highly reliable FC and behavioral measures. For the behavioral measure, we used the NIH Toolbox Cognition Total Composite Score (CogTotalComp\_Unadj), the only unadjusted behavioral summary measure in the HCP exhibiting excellent ICC (ICC = 0.935), based on repeated measures across the main HCP and HCP Retest datasets (see Figure \ref{fig:ICC_demo}). ICC is considered excellent if it exceeds 0.9 \citep{koo2016guideline}.  

To maximize reliability of FC, we averaged the estimated FC across all four runs for each participant, for a total of approximately $T=60$ minutes, and applied stringent censoring. Any participants missing at least one rs-fMRI run or missing the behavioral measure were excluded from the BWAS analysis, resulting in a sample size of $N=991$. The ground truth BWAS correlation $\rho^*_j$ at each edge $j$, while high-quality due to relatively high reliability of FC and behavior, will still be attenuated due to imperfect reliability of the FC and behavioral variable. Its expected value is $\mathbb{E}(\rho^*_j)=\rho_j\sqrt{R(x_{ij}^{(60)})R(y_i)} < \rho_j$, where $\rho_j$ is the true correlation value. Thus, performed bias correction as
$$
\tilde\rho_j^* = \frac{\rho^*_j}{\sqrt{R(x_{ij}^{(60)})R(y_i)}},
$$
which has expected value equal to the truth $\rho_j$. $R(y_i)=0.935$ as stated above, based on test-retest measures across the main and retest HCP datasets. For $R(x_{ij}^{(60)})$, we used the full HCP sample for higher accuracy, given the large number of edges to estimate. Since we do not have access to repeated 60-minute FC estimates from the full HCP sample, we extrapolated based on test-retest 30-minute FC estimate, as follows.  

First, we estimated the signal and noise variance for scan duration $T=30$, using test-retest FC estimates $x_{ij1}^{(30)}$ and $x_{ij2}^{(30)}$ from visits 1 and 2 of the main HCP.  Consider the measurement error model presented in Section \ref{methods:baseline_var}, where the estimated FC $x_{ij}^{(T)}$ is measured with error relative to the truth, $w_{ij}$. For a given edge $j$, assume that the errors $e_{ijk}^{(T)}$ across participants $i$ and visits $k$ are mutually independent and are independent of the true FC $w_{ij}$.  It is straightforward to show that the noise variance $\Var(e_{ij}^{(30)})=\frac{1}{2}\Var\big(x_{ij2}^{(30)} - x_{ij1}^{(30)}\big)$, and the signal variance $\Var\big(w_{ij}\big) = \Var\big(x_{ij}^{(30)}\big) - \Var(e_{ij}^{(30)})$, where the total variance $\Var\big(x_{ij}^{(30)}\big)$ is estimated as the mean over the visit-specific total variance estimates, $\Var\big(x_{ij1}^{(30)}\big)$ and $\Var\big(x_{ij2}^{(30)}\big)$. Note that the signal variance does not depend on the scan duration $T$.  Given the signal and noise variance, the ICC of 30-minute FC estimates is given by
$$
R(x_{ij}^{(30)})=\frac{\Var\big(w_{ij}\big)}{\Var\big(w_{ij}\big) + \Var(e_{ij}^{(30)})}.
$$

To extrapolate from $T=30$ to $T=60$ minute scan duration, we only need to estimate $\Var(e_{ij}^{(60)})$, since the signal variance is unaffected.  We can assume that the noise variance $\Var(e_{ij}^{(60)})$ for $T=60$ equals half the noise variance $\Var(e_{ij}^{(30)})$ for $T=30$, since variance is assumed to be inversely related to scan duration (see Section \ref{methods:baseline_var}).  Thus, the ICC of FC based on scan durations $T=60$ is estimated as
$$
{R}(x_{ij}^{(60)})=\frac{\Var\big(w_{ij}\big)}{\Var\big(w_{ij}\big) + \Var(e_{ij}^{(30)})/2},
$$

Based on this estimate of the ICC of FC from $T=60$ minute scans, we obtain bias-corrected ground truth BWAS correlations at each edge, $\tilde\rho_j^*$ as explained above. Then, to examine the effect of scan duration and censoring on attenuation of BWAS estimates relative to this ground truth, we varied scan duration from $T=5$ minutes to $T=30$ minutes and applied censoring at different levels, from none to expanded.  For each censoring method and scan duration, we estimated the BWAS correlation based on the corresponding FC estimate (see Section \ref{methods:FC}).  For a given estimate of the BWAS correlation, $\hat\rho_j^{(T)}$, we computed the attenuation as $\hat\rho_j^{(T)}/\tilde\rho^*_j$.  This analysis was repeated for each visit in the main HCP, providing two estimates of the attenuation for a given censoring level, duration and edge, which were then averaged. This analysis was performed for 36P- and FIX-processed data.

\bibliographystyle{apalike}
\bibliography{Scrubbing.bib}

\newpage
\appendix

\renewcommand\thesection{\Alph{section}}
\renewcommand\thesubsection{\thesection.\arabic{subsection}}
\renewcommand\thetable{S.\arabic{table}}
\renewcommand\thefigure{S.\arabic{figure}}
\setcounter{figure}{0}
\setcounter{table}{0}

%Figure S1 - Illustration of expanded censoring

\begin{figure}
    \centering
    \includegraphics[width=1\linewidth, trim=0 0 0 1cm, clip]{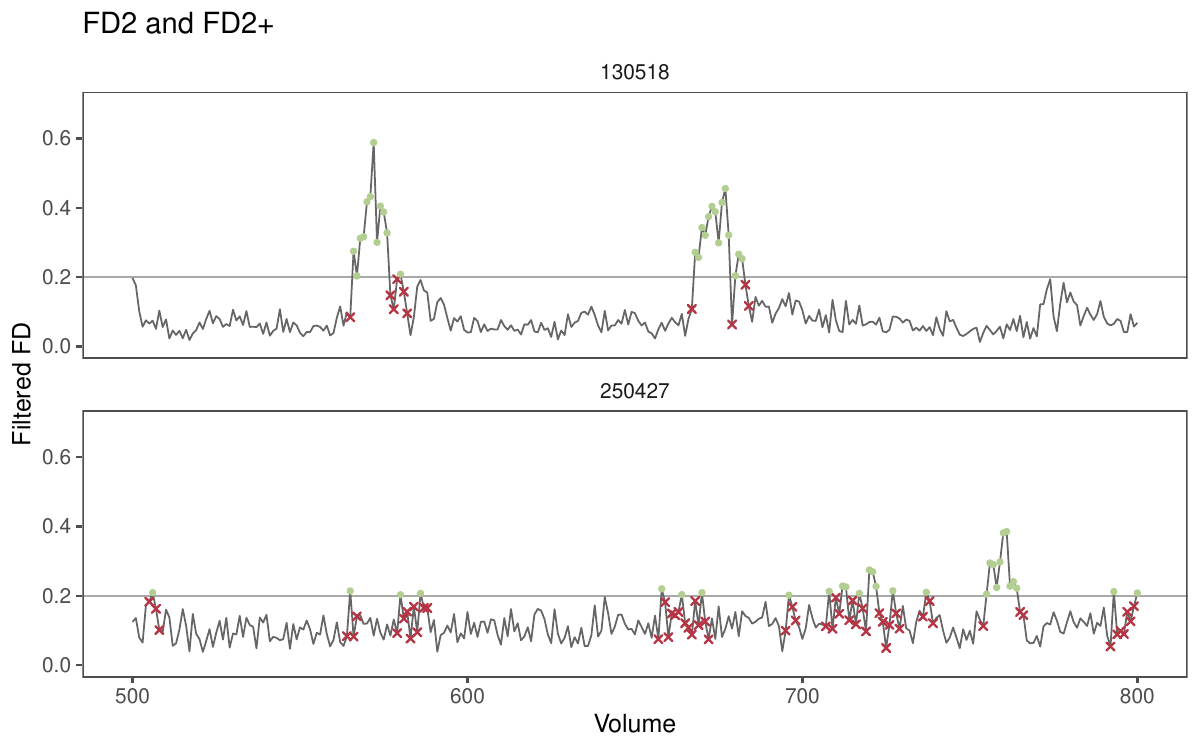}
    \caption{\small \textbf{Illustration of expanded censoring.} Green dots indicate volumes flagged with stringent censoring, and red $\times$'s indicate volumes additionally flagged with expanded censoring. The horizontal line represents FD = 0.2 mm, the cutoff for stringent censoring.     The three sessions displayed were chosen to represent a broad spectrum of expansion levels. Specifically, we show one session at the 5th percentile (participant 130518) and 95th percentile (participant 250427) of \textit{expansion}, the ratio between the number of volumes flagged with expanded censoring and with stringent censoring. Thus, participant 130518 shows an example of minimal expansion and participant 250427 shows an example of a high rate of expansion.}
    \label{fig:FD_Comparison}
\end{figure}

%Figure S2 - Parcellation

\begin{figure}
    \centering
    \includegraphics[width=0.9\linewidth]{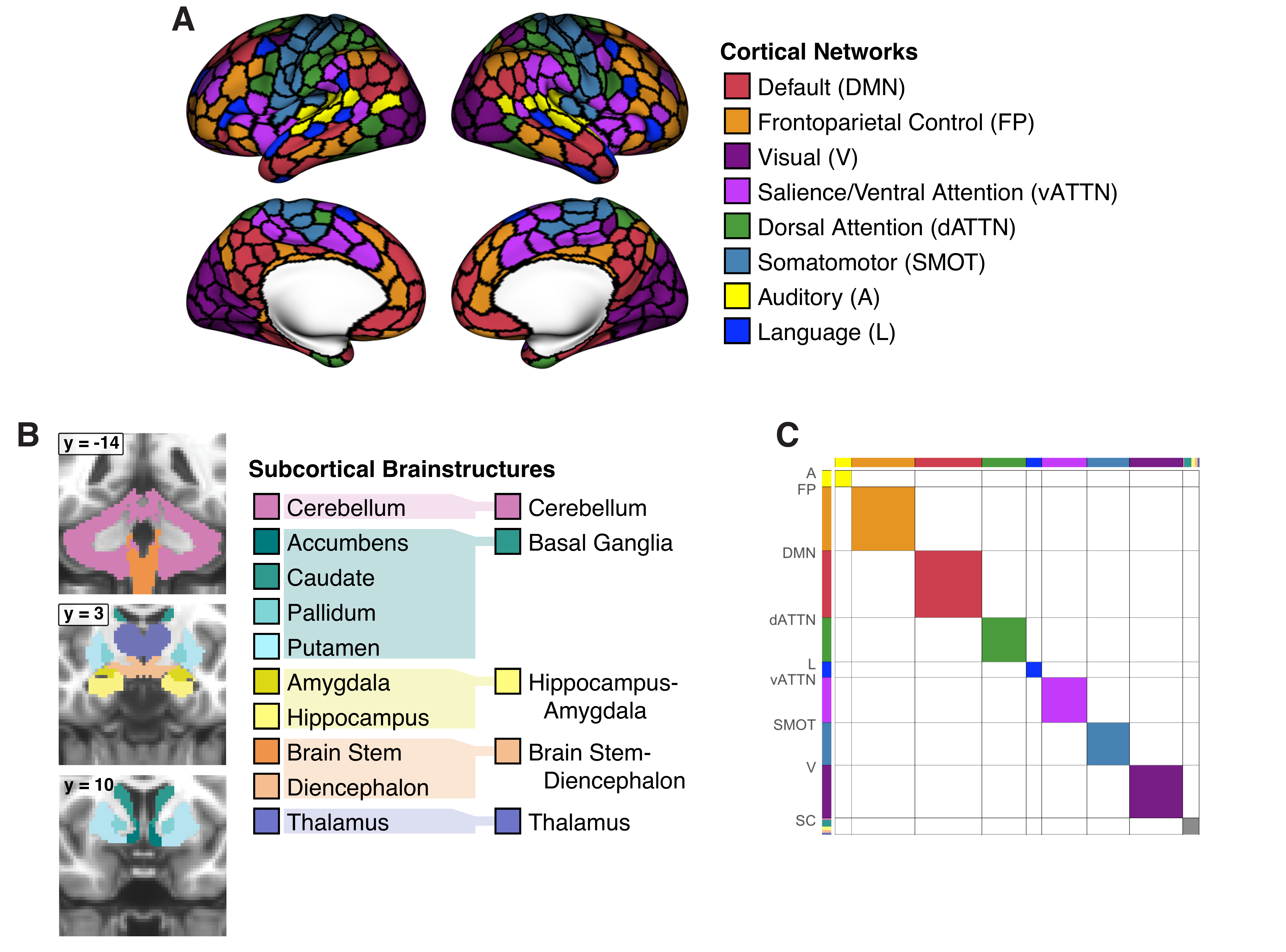}
    \caption{\small Cortical parcellation (A) and subcortical structures (B) used as the nodes in functional connectivity (C).}
    \label{fig:region_labels}
\end{figure}

%Figure S3 - Significance testing of FC error

\begin{figure}
    \centering
    {\Large \sffamily \qquad \underline{36P}} \\
    \includegraphics[scale=0.6, trim = 0 0 0 8mm, clip]{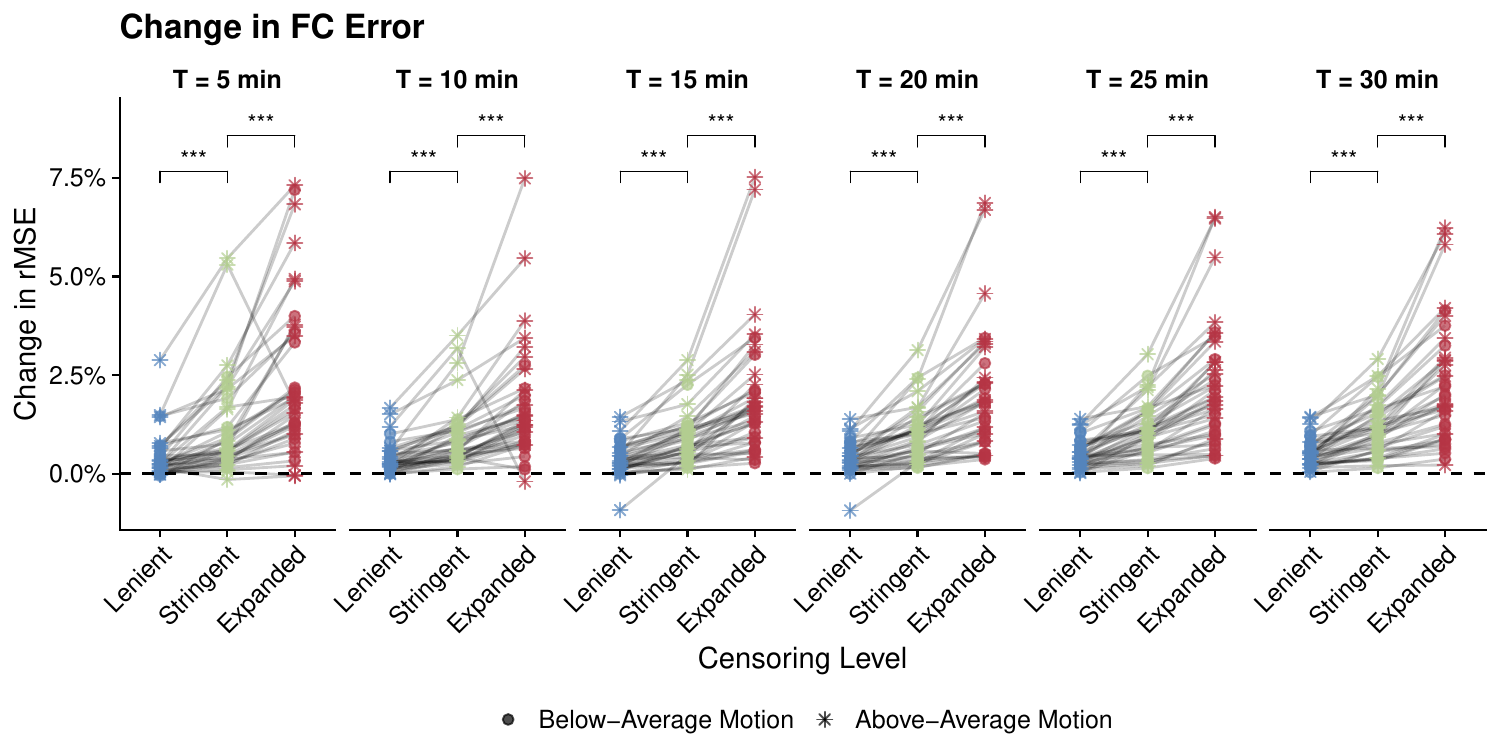} \\[12pt]
    {\Large \sffamily \qquad \underline{FIX}} \\
    \includegraphics[scale=0.6, trim = 0 0 0 8mm, clip]{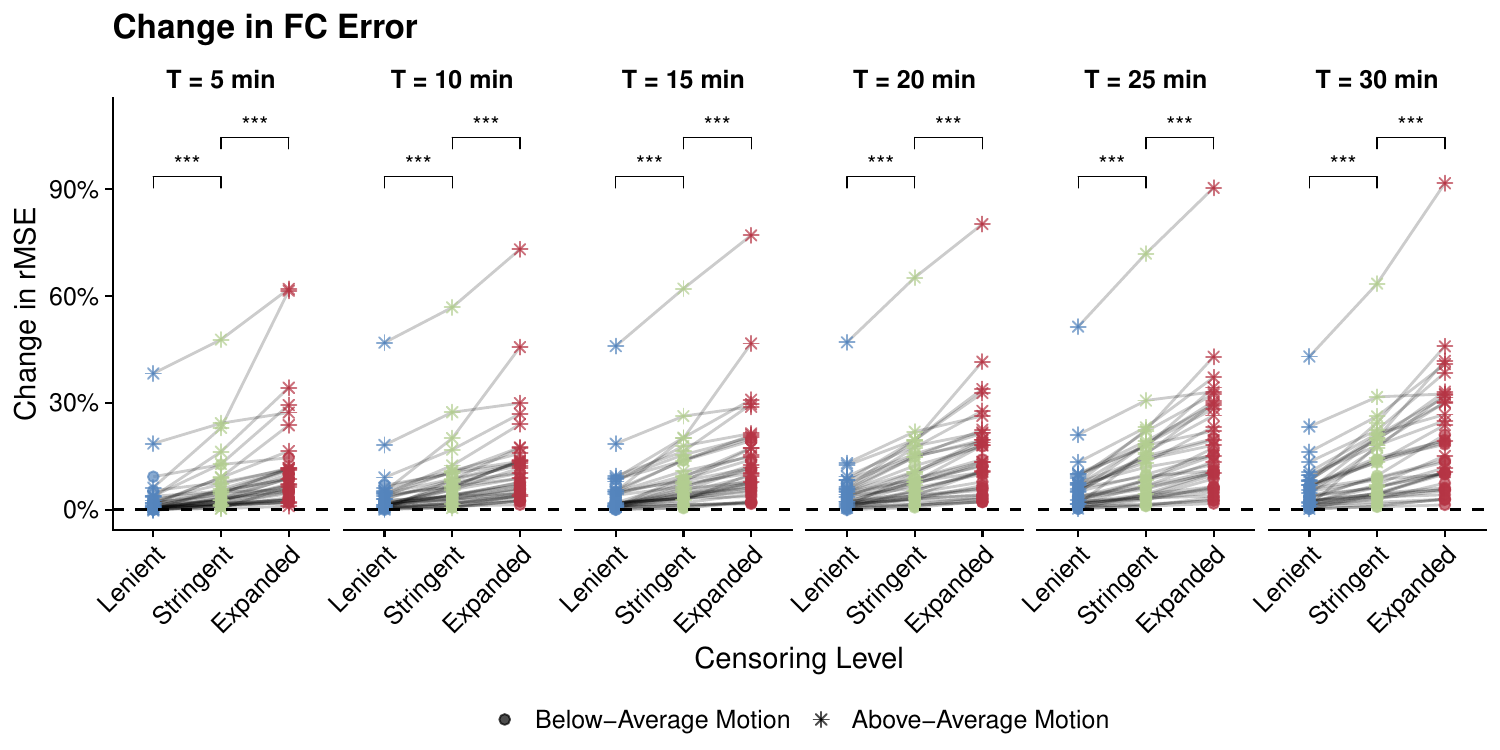}
\caption{\small \textbf{More aggressive censoring causes statistically significant worsening of FC error for all scan duration.} Stars indicate significance levels of comparisons surviving Bonferroni correction across all 18 tests (3 comparisons $\times$ 6 durations), based on paired Wilcoxon rank sum tests. All comparisons were found to be statistically significant with $p < 0.001$ after Bonferroni correction, indicating that more aggressive censoring (none to lenient (results not shown), lenient to stringent, or stringent to expanded) results in a statistically significant increase in FC error.}
\label{fig:tests_alldurations}
\end{figure}

%Figure S4 - Motion classification

\begin{figure}
    \centering
    \includegraphics[page=1, width=5in, trim = 0 33mm 0 1mm, clip]{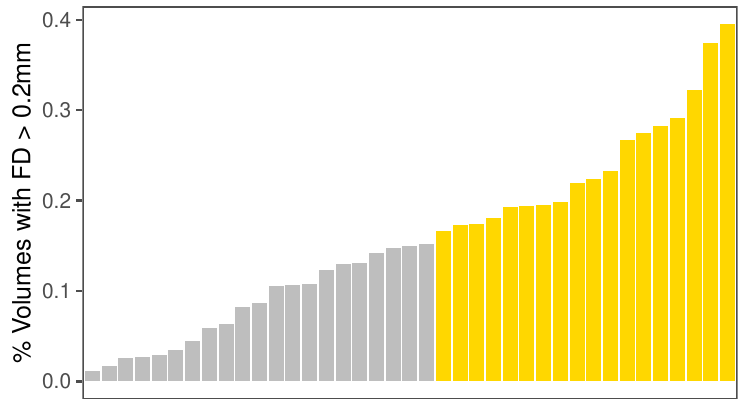} 
    \includegraphics[page=2, width=5in, trim = 0 0 0 1mm, clip]{plots/MotionSplit.pdf}
    \caption{\small \textbf{Classification of participants based on motion levels.} Participants are classified into below- and above-average motion groups based on percentage of high-motion volumes (FD $>$ 0.2mm).  Mean FD, shown in the bottom plot, was not used for classification since it is a measure of the magnitude but not the frequency of motion and is less effective at distinguishing participants. In both plots, participants are sorted by their percentage of high-motion volumes.}
    \label{fig:motion_split}
\end{figure}

%Figure S5 - Illustration of change in required duration analysis

\begin{figure}
    \centering
    \begin{tabular}{cc}
       By Edges  &  By Participants \\
       \includegraphics[width=0.45\textwidth]{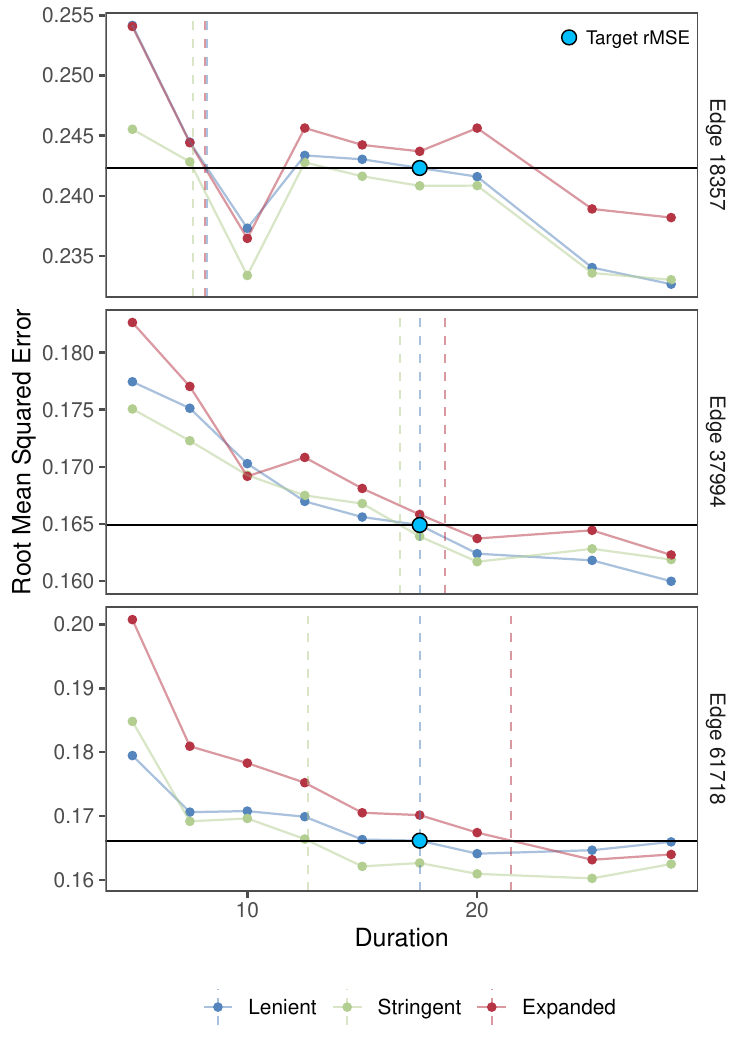}  & 
       \includegraphics[width=0.45\textwidth]{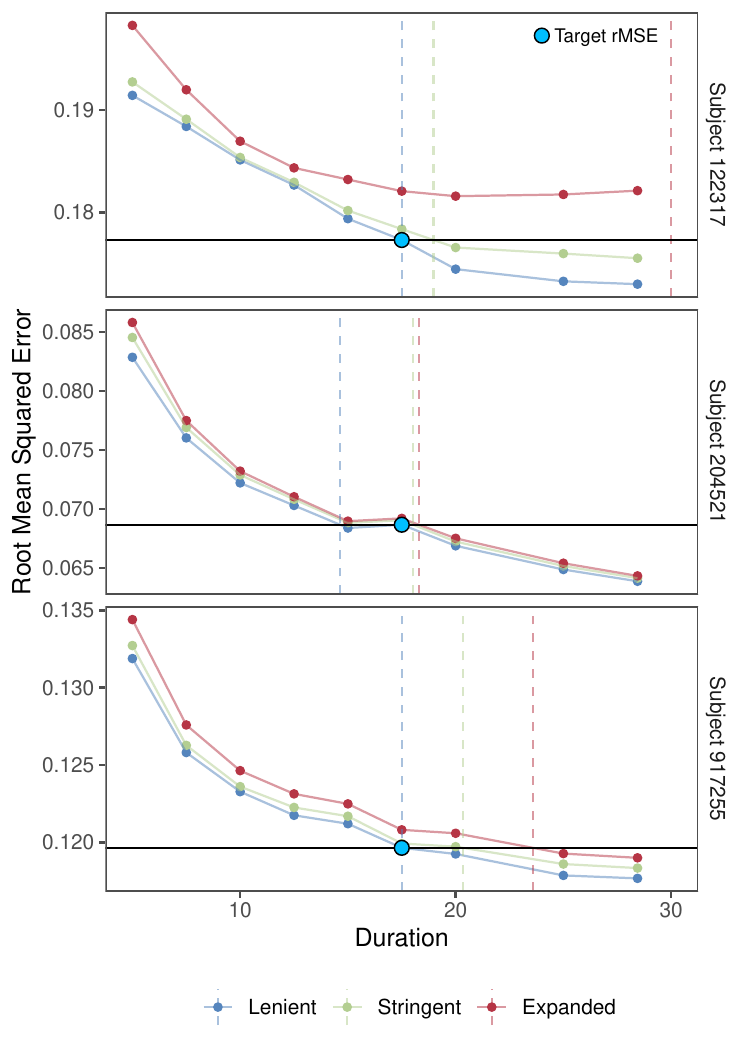} \\
    \end{tabular}
    \caption{\small \textbf{Illustration of change in required duration analysis.} The target FC accuracy (root mean squared error; rMSE) was set based on lenient censoring and duration of $T=17.5$ minutes, the midpoint of the range of durations considered ($5$ to nearly $30$ minutes). For each censoring level, the minimum duration required to achieve the target rMSE is indicated by a vertical dashed line, using linear interpolation to ``connect the dots'' between observed data points. This is the \textit{required scan duration} for a given censoring level and participant or edge. For a given censoring level, if the target rMSE is never achieved, then required scan duration is set to $30$ minutes (slightly above the observed maximum of $28.44$ minutes); if the target rMSE is above the actual rMSE across all scan durations, the required scan duration is set to $4$ minutes (slightly below the observed minimum of $5$ minutes).}
    \label{fig:durChange_illustration}
\end{figure}

%Figure S6 - Validation of repeated measures QC-FC approach

\begin{figure}
    \centering
    \begin{tabular}{cccc}
    \multicolumn{4}{c}{\underline{\textbf{Repeated Measures QC-FC: Motion Artifact/State}}} \\[10pt]
    \quad No Censoring & \quad Lenient & \quad Stringent & \quad Expanded \\
    \includegraphics[scale=0.18, trim=0 2cm 5mm 6mm, clip]{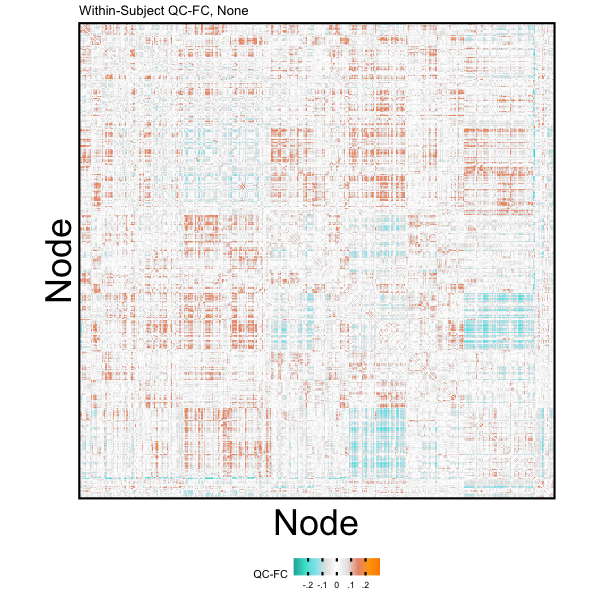} &
    \includegraphics[scale=0.18, trim=0 2cm 5mm 6mm, clip]{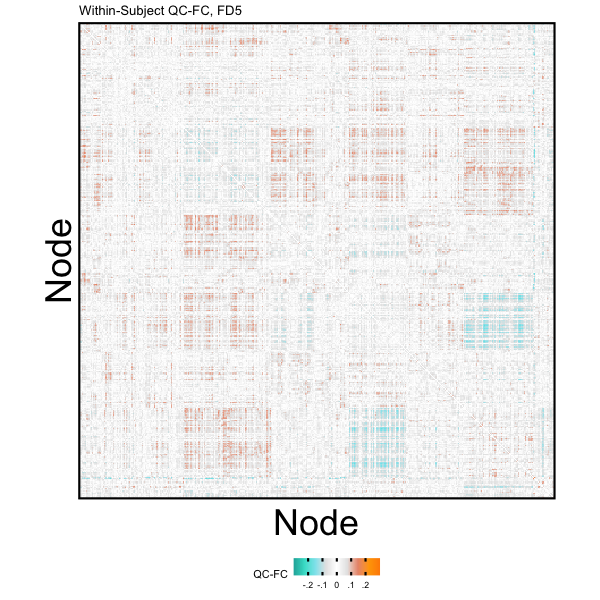} &
    \includegraphics[scale=0.18, trim=0 2cm 5mm 6mm, clip]{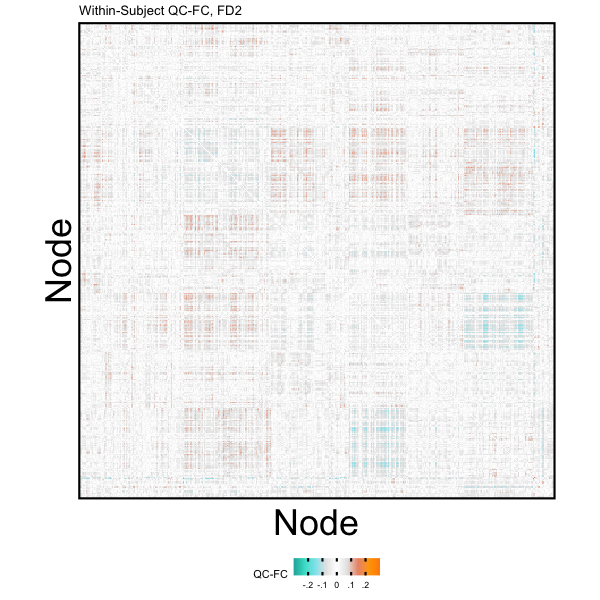} &
    \includegraphics[scale=0.18, trim=0 2cm 5mm 6mm, clip]{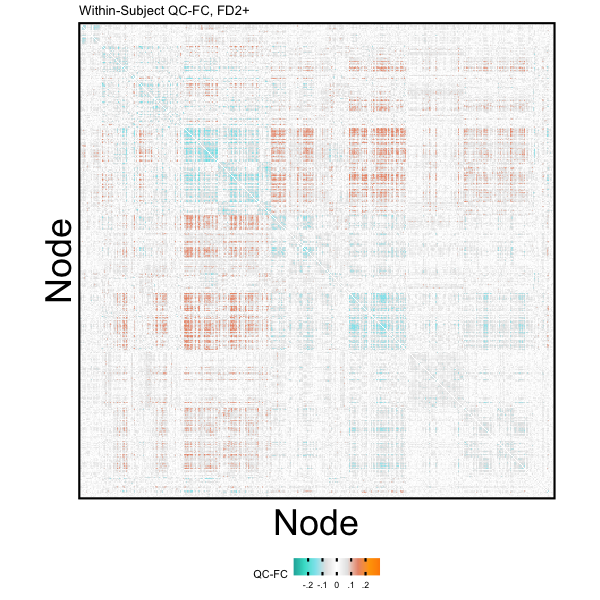} \\  
    \multicolumn{4}{c}{\underline{\textbf{Validation Version of Motion Artifact/State}}} \\[6pt]
    \quad No Censoring & \quad Lenient & \quad Stringent & \quad Expanded \\
    \includegraphics[scale=0.18, trim=0 2cm 5mm 6mm, clip]{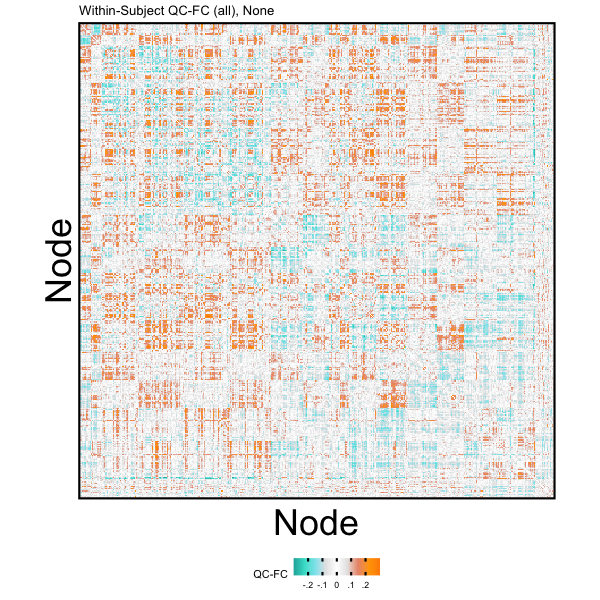} &
    \includegraphics[scale=0.18, trim=0 2cm 5mm 6mm, clip]{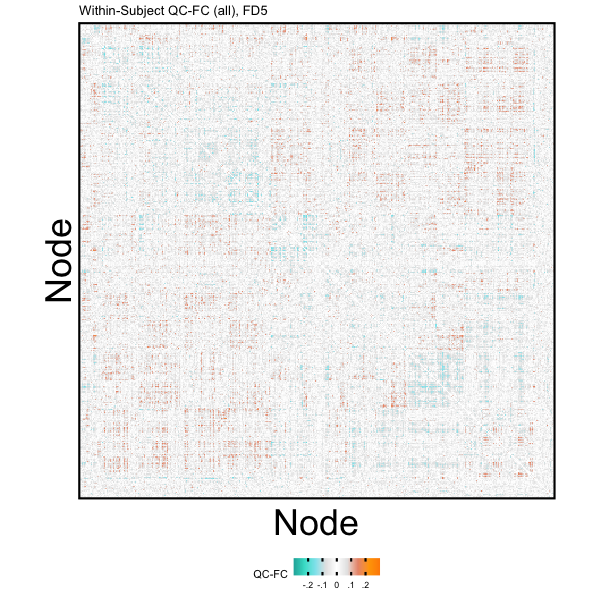} &
    \includegraphics[scale=0.18, trim=0 2cm 5mm 6mm, clip]{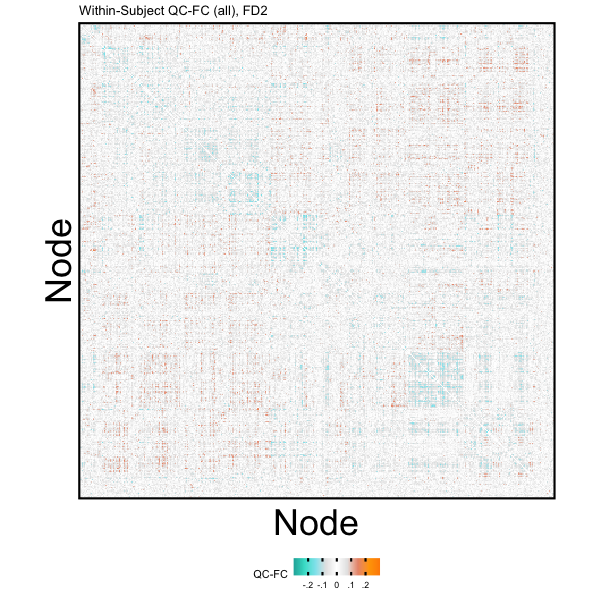} &
    \includegraphics[scale=0.18, trim=0 2cm 5mm 6mm, clip]{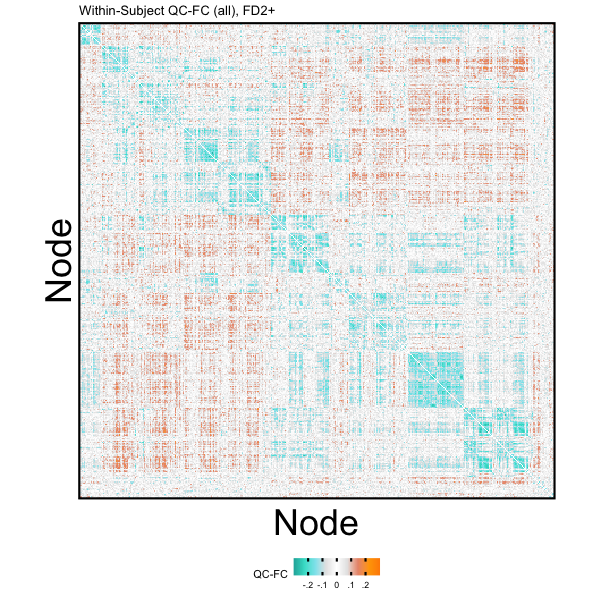} \\[10pt] 
    \end{tabular}
    \includegraphics[height=4mm]{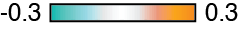} \\
    \caption{\small \textbf{Validation of repeated measures QC-FC approach.} The second row shows the validation version of the within-participant component of repeated measures QC-FC, computed as the correlation between changes in FC and changes in FD across high and low motion sessions from the same participants (Methods Section \ref{methods:QCFC}). Note that some differences are expected, since repeated measures QC-FC controls for between-participant effects, while the validation version does not. Yet both are consistent in showing that within-participant motion-FC associations are effectively mitigated by non-expanded censoring, but that expanded censoring has odd effects that are not consistent with removal of motion artifact alone. These effects of expanded censoring include inducing negative motion-FC associations within most networks (diagonal blocks) and between certain pairs of networks (off-diagonal blocks) and increasing motion-FC associations between other pairs of networks.)\\[20pt]}
    \label{fig:QCFC_36P_validation}
\end{figure}

% Figure S7 - ICC of behavioral variables in the HCP

\begin{figure}
    \centering
    \includegraphics[width=0.8\linewidth]{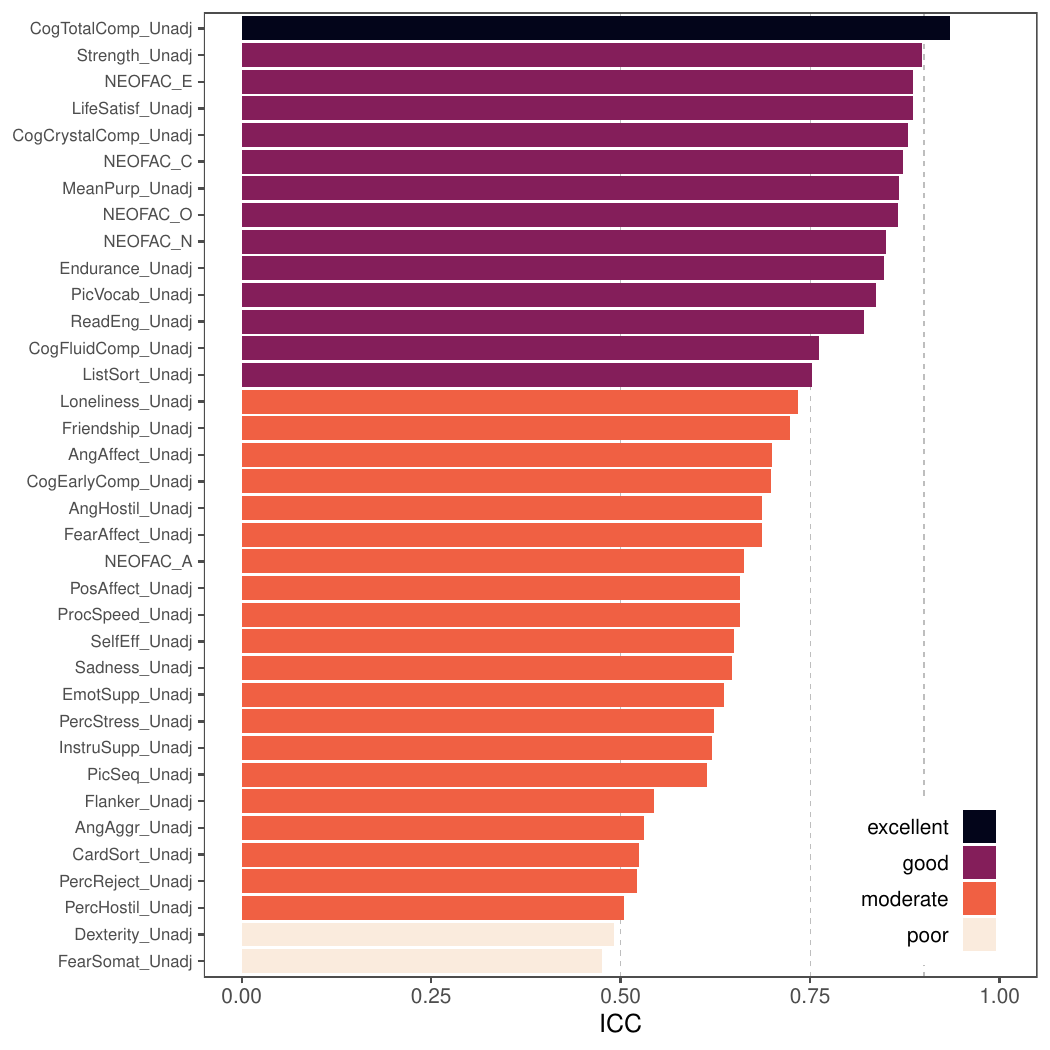}
    \caption{\small \textbf{Reliability of behavioral variables in the HCP.} Low reliability of behavioral variables and brain measures used in BWAS leads to attenuated BWAS correlations. Intra-class correlation coefficient (ICC) was determined based on repeated measures available across the main and retest samples of the HCP.  Continuous, summary variables in the Cognition, Emotion, Motor and Personality categories were included. Using standard cutoffs for ICC, most variables have moderate-good reliability. Only one variable (total cognition, CogTotalComp\_Unadj) has excellent reliability (ICC = 0.935), defined as ICC $\geq 0.9$ \citep{koo2016guideline}  For our empirical evaluation of the effect of imperfect FC reliability on BWAS attenuation, we use total cognition to minimize the effect of low behavioral reliability.}
    \label{fig:ICC_demo}
\end{figure}

% Figure S8 - BWAS variance inflated due to low FC reliability

\begin{figure}
    \centering
    \includegraphics[width=0.8\linewidth, trim = -2cm 0 0 0, clip]{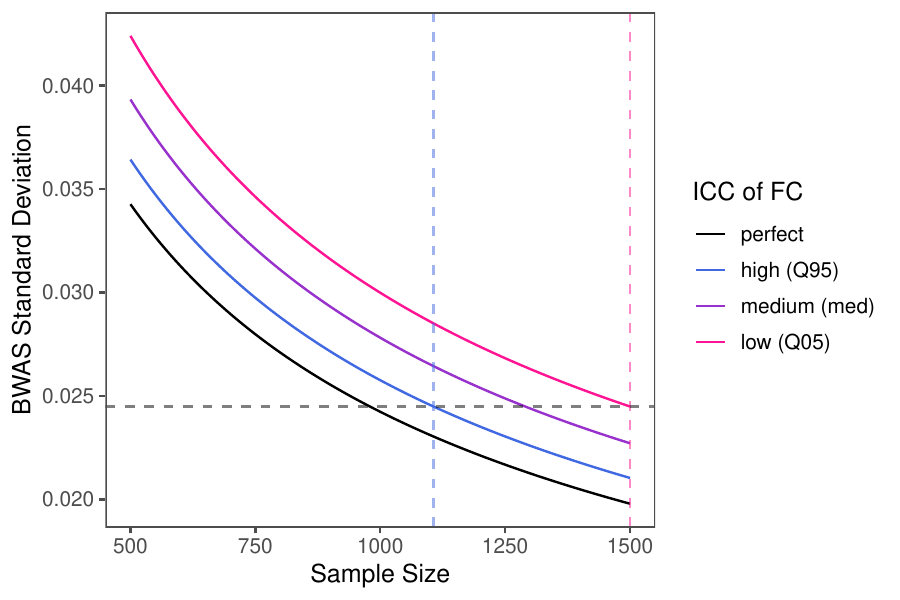}
    \caption{\small \textbf{BWAS variance is inflated due to low FC reliability.}  BWAS variance curves based on a hypothetical true correlation of $\rho_j = 0.5$ and empirical FC ICC values, based on scan duration $T=30$ min and stringent censoring in 36P-processed data. High, medium and low ICC values correspond to the 95th, 50th and 5th quantiles across edges, equal to 0.23, 0.52, and 0.80, respectively.  ICC of behavior was set to 0.934, the ICC the maximally reliable behavioral variable in the HCP (see Figure \ref{fig:ICC_demo}). Vertical lines illustrate that a smaller sample size ($n = 1,107$) can achieve the same BWAS variance when FC reliability is high, compared with when FC reliability is low with a larger sample size ($n = 1,500$).}
    \label{fig:BWAS_var_math}
\end{figure}

% Figure S9 - Bias correction of ground truth BWAS correlations

\begin{figure}
    \centering
    \begin{tabular}{cc}
       \hspace{1cm} {\Large 36P}  & \hspace{-1cm}{\Large FIX }  \\[10pt]
       \includegraphics[page = 1, height=2.7in, trim = 0 0 1in 8mm, clip]{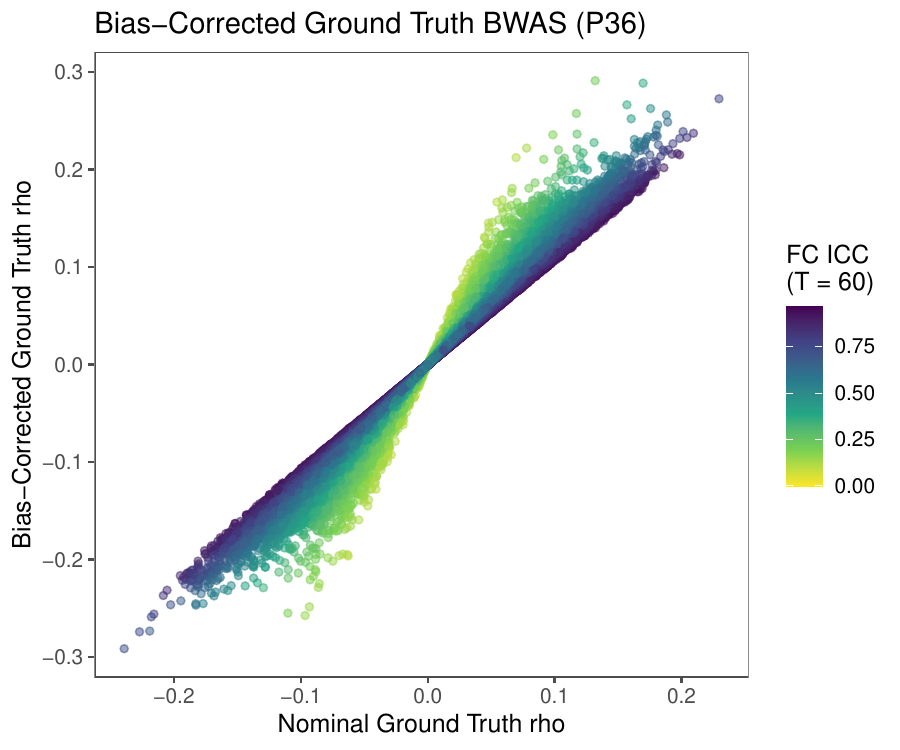}  & 
       \includegraphics[page = 2, height=2.7in, trim = 6mm 0 0 8mm, clip]{plots/BWAS/corrected_groundTruth.pdf}
    \end{tabular}
    \caption{\small \textbf{Bias correction of ground truth BWAS correlations.} The multiplicative bias correction factor is computed as the inverse of $\sqrt{R(x_{j})R(y)}$, where $R(x_{ij})$ is the ICC of FC at edge $j$, and $R(y)$ is the ICC of the behavioral measure (total cognition, ICC $\approx 0.93$). ICC of FC is computed for $T=60$ min scan duration, extrapolated from the signal and noise variance estimated from $T = 30$ minute scans. We exclude edges with ICC $< 0.1$ to avoid extreme correction ($0.26\%$ of edges for P36; $0.18\%$ of edges for ICA-FIX). The resulting correction ranges from approximately 1.05 for the most reliable edges (ICC $\approx 0.95$) to 3.27 for the least reliable edges (ICC $= 0.1$). The x-axis shows the nominal BWAS correlations between the ``ground truth'' FC at each edge (based on $T = 60$ minute scan duration with stringent censoring) and total cognition across nearly $1000$ participants in the main HCP. The y-axis shows the corrected BWAS correlations for each edge.}
    \label{fig:BWAS_groundTruth_correction}
\end{figure}

\end{document}